\documentclass[universe,review,accept,pdftex,moreauthors]{Definitions/mdpi}

\firstpage{1} 
\pubvolume{8}
\issuenum{11}
\articlenumber{556}
\pubyear{2022}
\copyrightyear{2022}
\externaleditor{Academic Editor: Arundhati Dasgupta and Alfredo Iorio}
\datereceived{9 September 2022} 
\dateaccepted{19 October 2022} 
\datepublished{25 October 2022} 
\DeclareMathOperator{\arctanh}{arctanh}

\hreflink{https://doi.org/10.3390/universe8110556} 
\pdfoutput=1

\usepackage{environ}
\NewEnviron{myequation}{%
\begin{equation}
\scalebox{0.95}{$\BODY$}
\end{equation}}

\Title{WKB Approaches to Restore Time in Quantum Cosmology: Predictions and Shortcomings}

\TitleCitation{WKB Approaches to Restore Time in Quantum Cosmology: Predictions and Shortcomings}

\newcommand{\ord}[1]{\mathcal{O} (#1)}
\newcommand{\dtau}[1]{\frac{\partial #1}{\partial \tau}}

\Author{Giulia Maniccia $^{1,2,}$*\orcidG{}, Mariaveronica De Angelis $^{3}$\orcidM{} and Giovanni Montani $^{4,1}$}

\AuthorNames{Giulia Maniccia, Mariaveronica De Angelis and Giovanni Montani}
\AuthorCitation{Maniccia, G.; De Angelis, M.; Montani, G.}

\address{$^{1}$ \quad Physics Department, “La Sapienza” University of Rome, P.le A. Moro 5, 00185 Roma, Italy; giovanni.montani@enea.it

$^{2}$ \quad INFN Section of Rome, “La Sapienza” University of Rome, P.le A. Moro 5, 00185 Roma, Italy

$^{3}$ \quad School of Mathematics and Statistics, University of Sheffield, Hounsfield Road, Sheffield S3 7RH, UK; mdeangelis1@sheffield.ac.uk

$^{4}$ \quad ENEA, FNS Department, C.R. Frascati, Via E. Fermi 45, Frascati 00044 (RM), Italy
}

\corres{Correspondence: giulia.maniccia@roma1.infn.it }

\abstract{
In this review, we analyse different aspects concerning the possibility to separate a gravity-matter system into a part which lives close to a quasi-classical state and a \lq \lq small” quantum subset. The considered approaches are all relying on a WKB expansion of the dynamics by an order parameter and the natural arena consists of the Bianchi universe minisuperspace. We first discuss how, limiting the WKB expansion to the first order of approximation, it is possible to recover for the quantum subsystem a Schrödinger equation, as written on the classical gravitational background. Then, after having tested the validity of the approximation scheme for the Bianchi I model, we give some applications for the quantum subsystem in the so-called \lq \lq corner” configuration of the Bianchi IX model. We individualize the quantum variable in the small one of the two anisotropy degrees of freedom. 
The most surprising result is the possibility to obtain a non-singular Bianchi IX cosmology when
the scenario is extrapolated backwards in time. In this respect, we provide some basic hints on the 
extension of this result to the generic cosmological solution. In the last part of the review, we consider the same scheme to the next order of approximation identifying the quantum subset as made of matter variables only. This way, we are considering the very fundamental problem of non-unitary morphology of the quantum gravity corrections to quantum field theory discussing some proposed reformulations. Instead of constructing the time dependence via that one of the classical gravitational variables on the label time as in previous works, we analyse a recent proposal to construct time by fixing a reference frame. This scheme can be reached both introducing the so-called “kinematical action”, as well as by the well-known Kuchar--Torre formulation. In both cases, the Schr\"odinger equation, amended for quantum gravity corrections, has the same morphology and we provide a cosmological implementation of the model, to elucidate its possible predictions.
}


\begin{document}

\section{Introduction}

All the canonical formulations for quantum gravity \cite{bib:dewitt1-1967,bib:dewitt2-1967,bib:dewitt3-1967,bib:kuchar-1981,bib:thiemann-2006,bib:cianfrani-canonicalQuantumGravity} have to deal with two fundamental questions: one concerning the construction of a suitable time variable for the dynamics, and the other one on the determination of a correct classical limit coinciding with General Relativity (GR).

The first question has been widely addressed in the literature \cite{bib:rovelli-1991-time,bib:kuchar-torre-1991,bib:brown-kuchar-1995,bib:montani-2002,bib:mercuri-montani-2004-dualism,bib:montani-zonetti-2008,bib:montani-castellana-2008,bib:cianfrani-montani-zonetti-2009-brown}  and the most commonly accepted idea is that the dynamics must be described via the introduction of a \lq\lq relational time'' \cite{bib:rovelli-1991-time}. The second point on the classical limit is a rather natural question for the metric approach, related to the Wheeler--DeWitt (WDW) equation, but it becomes a puzzling question in Loop Quantum Gravity \cite{bib:montani-cianfrani-2008-gr,bib:thiemann-book}.
However, these two points are unavoidably related to the possibility to reconstruct an evolutionary quantum field theory when, starting from a purely quantum gravity approach in the presence of matter, we consider the classical limit on the geometrical component only. This theme contains also the challenging perspective to determine quantum gravity corrections to quantum field~theory.

The first well-known attempt to reconstruct a Schr\"odinger functional theory for quantum matter from Canonical Quantum Gravity was performed in \cite{bib:vilenkin-1989}, limiting attention to a minisuperspace model, in which a \lq \lq small'' quantum subset (not necessarily restricted to matter) is recovered on a quasi-classical geometrodynamics. {For previous approaches where gravity was treated on a classical background level, see}  \cite{bib:lapchinsky-1979,bib:banks-1985}.
This line of research was then {expanded} in \cite{bib:kiefer-1991}, where a full development of the gravity-matter dynamics has been performed in terms of a single order parameter, combining the Planck length and the Newton constant {(see also} \cite{bib:singh-1990} {where such expansion was considered but not fully developed)}.
As shown in \cite{bib:montani-digioia-maniccia-2021}, the latter analysis is very similar to that in \cite{bib:vilenkin-1989}, but it is extrapolated to the next order where quantum gravity corrections to quantum field theory must appear.
However, constructing a time variable via the time dependence of the classical limit of the metric has been recognized as affected by a non-trivial shortcoming, {i.e.,} the emergence of non-unitarity features in the Schr\"odinger equation.

Here, we review the original formulations giving some sample of the implementation
to the idea in \cite{bib:vilenkin-1989} to specific cosmological situations, with particular reference to the \lq \lq corner configuation'' of a Bianchi IX dynamics. We also discuss the comparison of the Wentzel--Kramer--Brillouin (WKB) analysis to the standard Arnowitt--Deser--Misner (ADM) reduction \cite{bib:arnowitt-1962} of the dynamics for the Bianchi I model, confirming that a basic assumption of the proposed formulation is the
\lq \lq smallness'' of the quantum phase space available to the~subsystem.

Then, in the second part of the review, the question concerning the non-unitarity problem is addressed in more detail and proposals for its solutions \cite{bib:kiefer-2018,bib:bertoni-venturi-1996,bib:montani-digioia-maniccia-2021,bib:maniccia-montani-2022} are presented and discussed.
In particular, we focus our attention to a cosmological implementation of the idea developed in \cite{bib:maniccia-montani-2022} that the time variable can be constructed \emph{a la} Kuchar--Torre \cite{bib:kuchar-torre-1991}, {i.e.,} fixing a Gaussian reference frame, which is \lq \lq materialized'' as a fluid in the dynamics. In this framework, both the violation of the so-called strong energy condition and the non-unitarity of the quantum corrections to quantum field theory are simultaneously overcome.

The main aim of the present review is to focus attention to a theory that lives between quantum gravity and quantum field theory on curved space-time, {i.e.,} the co-existence of quantum field theory for matter with weak quantum features of the background gravitational field.
A convincing solution to the non-unitarity problem is therefore a central theme in this perspective and we provide a valuable picture on both the existing problems and the most promising formulations.

The review is structured as follows. In Section \ref{sec:formalism}, we illustrate the general formalism of the minisuperspace reduction of a gravity-matter system, addressing the problem of time concerning the cosmological wave function. In Section \ref{sec:vilenkin}, we present the work \cite{bib:vilenkin-1989} that proposes a solution by a semiclassical separation of the system, with a brief discussion on the boundary conditions for such wave function in Section \ref{ssec:boundary}. In Section \ref{sec:appl-vilenkin-bianchiI} we show the implications of such model for a Bianchi I universe. Section \ref{sec:appl-vilenkin-bianchiIX} presents instead the results of the model for the Bianchi IX universe, considering the vacuum case (Section~\ref{ssec:bianchi-IX-vacuum}), the presence of a cosmological constant and scalar field (Section \ref{ssec:bianchi-IX-cosmconst}), the Taub model (Section~\ref{ssec:taub}) and the generic inhomogenenous extension (Section \ref{ssec:bianchi-IX-ihnom}). In Section \ref{sec:WKB} we present the Wentzel--Kramer--Brillouin expansion, whose special case is \cite{bib:vilenkin-1989}, discussing in Section \ref{ssec:non-unitarity} the proposal \cite{bib:kiefer-1991} that uses such a method to compute quantum gravity corrections to the matter sector dynamics and the following non-unitarity issue, while in Section \ref{ssec:BO} we review the Born--Oppenheimer scheme proposed in \cite{bib:bertoni-venturi-1996} discussing its shortcomings.
Section \ref{sec:kin-fluid-unitarity} contains the recent proposals to solve both the problems of time and non-unitarity by implementing as a clock either the kinematical action (Section \ref{ssec:kin-action}) or the reference frame fixing procedure (Section \ref{ssec:gaussian-ref-frame}), presenting a cosmological implementation of the latter in Section \ref{ssec:fluid-applic}. Further discussion and conclusions are provided in Section \ref{sec:conclusions}.

\section{General Formalism: The Minisuperspace Analysis}\label{sec:formalism}

Let us preliminarily fix the general context in which we will develop our analysis. In this respect, we consider a minisuperspace cosmological model~\cite{bib:misner-gravitation,bib:montani-primordialcosmology, bib:capozziello-lambiase-2000}, namely a reduction of the Wheeler superspace in presence of symmetries, with the line element in the Arnowitt--Deser--Misner (ADM) formulation \cite{bib:arnowitt-deser-misner-1960} such as:
\begin{linenomath}
\begin{equation}
	ds^2 = N^2(t) dt^2 - h_{ab}\, \sigma^a \sigma^b
	\, ,
	\label{eq:ADMline}
\end{equation}
\end{linenomath}
where $h_{ab}$ ($a,b=1,2,3$) is a function of $n$ time dependent variables $g^a$, the 1-forms $\sigma^a$ define the specific isometry of the considered model, {e.g.,} the Bianchi universes \cite{bib:montani-battisti-imponente-2008-mixmaster,bib:landau2} and $N$ is the lapse function, {whose} specification determines the adopted time variable. {An application of the above considerations can be developed for $f(R)$ gravity as analysed in}~\cite{bib:faraoni-capozziello-2010, bib:capozziello-bajardi-2022}.

In the Hamiltonian representation \cite{bib:misner-gravitation}, the action for the minisuperspace takes the general form 
\begin{linenomath}
\begin{equation}
	S_{MSS} = \int dt \left\{  
	p_a\,\dot{g}^a - NH_{MSS}\right\}
	\, , 
	\label{eq:action-minisuperspace}
\end{equation}
\end{linenomath}
$p_a$ being the conjugate momenta to the configurational variables $g^a$  ($a=1,2,...,n$), and the superHamiltonian reads
\begin{linenomath}
\begin{equation}
	H_{MSS}(g^a, p_a) = G^{ab}p_a p_b + V(g^a)
	\, .
	\label{eq:hamiltonian-minisuperspace}
\end{equation}
\end{linenomath}
Here $G^{ab}$ denotes the minisupermetric, encoding metric properties in the minisuperspace and in general having a pseudo-Riemmannian character, while  $V(g^a)$ is a potential term due to the spatial curvature of the considered cosmological model. 
Additional contributions to both of them can come from the introduction of matter in the dynamics. Of particular reference is, in this respect, the presence of a self-interacting scalar field $\phi$, interpretable as the inflaton and responsible for the inflationary phase of the Universe.
In such a case, the superHamiltonian becomes
\begin{linenomath}
\begin{equation}
	H_{MSS}(g^a,p_a) + \frac{1}{2\sqrt{h}}p_{\phi}^2 + \sqrt{h}\, U(\phi )
	\, , 
	\label{eq:hamiltonian-inflaton}
\end{equation}
\end{linenomath}
$p_{\phi}$ being the conjugate momentum to the scalar field, $U(\phi)$ its self-interaction potential and $h\equiv \det{h_{ij}}$. 

Clearly, by varying the action with respect to $N$, we get that the (total) superHamiltonian \eqref{eq:hamiltonian-inflaton} identically vanishes and this fact reflects the time diffeomorphism invariance of the theory. Thus, implementing the Dirac prescription \cite{bib:cianfrani-canonicalQuantumGravity} for the canonical quantization of a constrained system, we naturally arrive to the following Wheeler--DeWitt (WDW) equation
\begin{linenomath}
\begin{equation}
	\left[ -\hbar^2G^{ab}(g^a)
	\frac{\partial \,}{\partial g^a}\frac{\partial \,}{\partial g^b}
	- \frac{\hbar^2}{2\sqrt{h}}\frac{\partial^2\,}{\partial \phi^2} + V(g^a) + \sqrt{h} \,U(\phi )\right] \psi = 0
	\, , 
	\label{eq:WDW}
\end{equation}
\end{linenomath}
where the Universe wave function $\psi (g^a,\phi )$ is intrinsically taken over $3$-geometries \cite{bib:misner-gravitation} since the spatial diffeomorphisms leave the 1-forms $\sigma^a$ invariant.
Above, we have chosen the so-called natural operator ordering, {i.e.,} the functions of $g^a$ are taken always on the left of the corresponding partial differentiations in constructing the quantum operator constraint. In this case, the minisupermetric is often redefined by a global scaling as $G^{ab}\rightarrow \sqrt{h} \,G^{ab}$, when the whole constraint is multiplied by $\sqrt{h}\neq 0$. 
Other operator orderings are available and classes of equivalence can be established \cite{bib:kuchar-1981}; in particular, we mention the choice of a symmetric superHamiltonian operator (for a justification see \cite{bib:lulli-cianfrani-montani-2012}). 

Equation \eqref{eq:WDW} is affected by the so-called \lq\lq frozen formalism'' problem, {i.e.,} no time evolution emerges in terms of the wave function dependence on an external time parameter, as will be discussed in Section~\ref{ssec:problemtime}. 
However, it is a well-known result \cite{bib:dewitt1-1967,bib:dewitt2-1967,bib:dewitt3-1967} that the WDW equation has a Klein--Gordon-like structure due to the pseudo-Riemmannian nature of $G^{ab}$. In fact, taking $h^{1/4}$ as a generalized coordinate, we easily see that it has a different signature with respect to the remaining ones, including also the scalar field. 
In the spirit of the relational approach proposed in \cite{bib:rovelli-1991-time}, see also \cite{bib:montani-primordialcosmology}, the scalar field can be taken as a matter clock, even though it has the same signature of the \lq \lq space-like'' variables in $G^{ab}$. 

Thus, the quantization of a minisuperspace model corresponding to a Bianchi Universe is reduced to the quantum dynamics of a relativistic particle \cite{bib:bjorken-drell-rqm} which is affected by a subtle question concerning the construction of a Hilbert space. 
In particular, the presence of the two potential terms in Equation~\eqref{eq:WDW} prevents, in many situations, the possibility for a frequency separation, which can be achieved under specific assumptions or in suitable asymptotic limits. In this respect, it is worth stressing that we consider here the WDW equation as a single particle dynamics \cite{bib:wald-1993}, see also \cite{bib:giovannetti-montani-2022}, without considering the so-called \lq\lq third quantization approach'' \cite{bib:caderni-1984,bib:mcguigan-1988}, that was first introduced as production of \lq\lq baby Universes'' in relation to the cosmological constant problem \cite{bib:hawking-1987,bib:giddins-1988,bib:coleman-1988,bib:rubakov-1988,bib:giddings-1989,bib:vilenkin-1994-rassegna}.
Finally, we observe that in Quantum Gravity, according to a very general prescription \cite{bib:isham-book-1993}, the choices of $h^{1/4}$ or of $\phi$ as internal time coordinates can be performed after or before the quantization procedure. In the former case, the quantization is covariantly performed, without specifying any explicit expression for the lapse function. 
Instead, in the latter case, the choice is performed on a classical level by fixing the temporal gauge which naturally leads to the ADM-reduction \cite{bib:arnowitt-1962} of the classical variational principle and therefore to a Schr\"{o}dinger-like quantum dynamics for the Universe wave function. 

\subsection{The Wave Function and the Problem of Time in Quantum Cosmology}\label{ssec:problemtime}
An important approach to quantum cosmology and its many applications regards the semiclassical approximation of the Universe. Indeed, in the full quantum picture, there is still some discussion regarding the probabilistic interpretation of the Universe wave function. This aspect is not straightforward since the wave function itself does not evolve in \lq \lq time'' due to the vanishing of the WDW Equation \eqref{eq:WDW} \cite{bib:cianfrani-canonicalQuantumGravity, bib:kiefer-sand-2022, bib:halliwell-1989(2), bib:wiltshire-1995}.
In the canonical quantum picture \cite{bib:dirac-lecturesOnQuantumMechanics}, this is equivalent to a timeless Schr\"{o}dinger equation with null eigenvalues describing a trivial evolution leading to the so-called \emph{problem of time}~\cite{bib:isham-book-1993,bib:mercuri-montani-2004-framefixing,bib:mercuri-montani-2004-dualism,bib:cianfrani-montani-zonetti-2009-brown,bib:kiefer-2013-review}. This issue has been long discussed in the literature since the formulation of the DeWitt theory \cite{bib:feinberg-1995,bib:kuchar2011-review,bib:bojowald-2018,bib:gielen-2021,bib:goltsev-2021,bib:peter-kiefer-2022,bib:altaie-beige-2022}{: for example, in }\cite{bib:bojowald-2018} {some time choices (scalar field, cosmological constant conjugate, and proper time) models are discussed via a semiclassical expansion in $\hbar$}. Indeed, the time coordinate could in principle be regarded together with the gravitational degrees of freedom and integrated over \cite{bib:vilenkin-1989}, such that there is no clear choice for the definition of another time parameter; subsequently, the definition of a conserved and well-defined probability distribution is troublesome, unless one imposes further conditions, {e.g.,} hermicity of the Hamiltonian \cite{bib:partouche-2021} or finiteness of the probability density \cite{bib:he-2015}. One of the most followed approaches is the definition of a relational time \cite{bib:rovelli-1990,bib:rovelli-1991-time,bib:rovelli-1991-quantrefsyst,bib:kuchar-torre-1991,bib:wald-1993,bib:feinberg-1995,bib:brown-kuchar-1995,bib:thiemann-2006} to recover a time parameter leading to a Schr\"{o}dinger dynamics; such \lq\lq emergence of time'' has been discussed not only for quantum gravity but also in the context of non-relativistic quantum mechanics, for example in \cite{bib:briggs-2015}.

This identification of a proper time-like variable avoiding the frozen formalism leads to different results whether it is tackled before or after quantization \cite{bib:guven-1992}. Hence, to give a meaningful probabilistic interpretation to the wave function of the Universe, one can pursue two different approaches. In the first, the super-Hamiltonian constraint is classically solved and then the resulting Schr\"odinger equation is quantized \cite{bib:misner-1969,bib:montani-benini-2006}, {i.e.,} the reduced phase space quantization (RPSQ) \cite{bib:henneaux-1992,bib:thiemann-2006-rpsq}. The RPSQ is the most straightforward method because it is an exact procedure requiring no WKB approximation based on the wave function of the Universe, even if its mass-like term is time-dependent and the Hamiltonian density is non-local. While in the second case, one implements both the WKB and Born--Oppenheimer (BO) approximation (see Sections~\ref{sec:WKB} and~\ref{ssec:BO}), that is essentially Vilenkin's approach (see also the discussion and application in Section~\ref{sec:appl-vilenkin-bianchiI}). 

DeWitt himself observed that \eqref{eq:WDW} is equivalent to a $n$-dimensional Klein--Gordon equation with variable mass term \cite{bib:dewitt1-1967,bib:dewitt2-1967,bib:dewitt3-1967} given by $-\sqrt{h}\, R^{(3)} $, being $R^{(3)}$ the scalar curvature associated to the induced metric $h_{ij}$, and so one could implement a Klein--Gordon-like inner product. However, being the mass term not necessarily positive, and the Hamiltonian containing second derivatives in the metric coordinates, such definition could give negative probabilities for the wave functional, {i.e.,} negative frequency components. This feature can be avoided in some special cases \cite{bib:wald-1993} with \emph{ad-hoc} conditions, but it remains standing in the general case, leaving some concerns on how to interpret the wave functional itself.

\section{Semiclassical and Quantum Universes: Vilenkin's Approach}\label{sec:vilenkin}

Vilenkin's proposal \cite{bib:vilenkin-1989} will be the starting point of our analysis, that aims to reconcile the WDW equation with a functional field theory formalism for gravity in the minisuperspace via a semiclassical expansion. To better show the feasibility of this model in the context of quantum cosmology, some key implementations to Bianchi universes are then examined in Sections~\ref{sec:appl-vilenkin-bianchiI} and \ref{sec:appl-vilenkin-bianchiIX}.

Starting from the interpretation of DeWitt and following the path of a relational time study, Vilenkin's work \cite{bib:vilenkin-1989} suggested to separate the Universe variables into semiclassical and quantum components.  This separation is indeed valid at some point, since gravity has a full quantum behaviour only near the Planck scale, and many physical phenomena relevant in cosmology happen at lower energies.

Vilenkin first considered the case in which the whole Universe behaves semiclassically. In the homogeneous minisuperspace setting, such a system can be described by a wave function of the form
\begin{linenomath}
\begin{equation}\label{eq:defVilenkinSemiclassical}
    \Psi(h) = A(h) e^{\frac{i}{\hbar} S(h)} \,,
\end{equation}
\end{linenomath}
where we label by $h^{a}$ all the semiclassical superspace variables (both gravity and matter fields) and $S(h^{a})$ the classical action that must be a real function, while $A(h)$ encodes the semiclassical features. The WDW equation reads as
\begin{linenomath}
\begin{equation}\label{eq:Vil-WDW-semicl}
    \left( -\hbar^2 \nabla^{(c) 2} + U^{(c)}\right) \Psi =0 \,,
\end{equation}
\end{linenomath}
being $U^{(c)} = \sqrt{h} \,U(\phi) + V(g^{a})$ the potential associated to all semiclassical variables and $\nabla^{(c)}_a$ the derivative with respect to $h^a$ (we are using the superscript $^{(c)}$ to identify the semiclassical components). Such a writing is superfluous since at this stage the whole Universe behaves semiclassically, but it will come into play later by considering the more general case. 
A perturbative expansion of $S$, and so of $\psi$, can be implemented in powers of the Planck constant due to the semiclassical feature of the Universe (in the original work, the expansion was performed in a parameter proportional to $\hbar$ and the $\hbar$ in \eqref{eq:defVilenkinSemiclassical} was absorbed inside the function $S$; here, for clarity, it is collected in front). This allows to study the dynamics going from the lowest order corresponding to the classical limit $\hbar \rightarrow 0$, to higher orders in such parameter. The procedure is clearly linked to the Wentzel--Kramer--Brillouin (WKB) approximation \cite{bib:dunham-1932} explained in Section~\ref{sec:WKB}, that uses an ansatz very similar to \eqref{eq:defVilenkinSemiclassical} but with a complex exponential and without the explicit separation of a semiclassical amplitude.
The expansion of \eqref{eq:Vil-WDW-semicl} brings at $\ord{\hbar^0}$
\begin{linenomath}
\begin{equation}\label{eq:Vilenkin-HJ}
    (\nabla^{(c)} S)^2 + U^{(c)} =0 \, ,
\end{equation}
\end{linenomath}
that is the Hamilton-Jacobi (HJ) equation for $S$, ensuring the classical limit of the model.
The next order $\ord{\hbar}$ gives
\begin{linenomath}
\begin{equation}\label{eq:VilenkinAmplitude}
    2 \nabla^{(c)} A \cdot \nabla^{(c)} S + A \, \nabla^{(c)2} S =0,
\end{equation}
\end{linenomath}
where the supermetric $G_{ab}$ is implicitly assumed by the scalar product symbol $(\cdot)$; this is equivalent to the conservation of the following current
\begin{linenomath}
\begin{equation}\label{eq:VilenkinCurrent}
    j^{(c)\,a} = |A|^2 \,\nabla^{(c)\,a} S \,,
\end{equation}
\end{linenomath}
whose interpretation can now be understood together with the associated semiclassical probability distribution $\rho^{(c)}$. Indeed, the action $S$ defines a congruence of classical trajectories, as follows from \eqref{eq:Vilenkin-HJ}; each point $h^{a}$ in a classically allowed region in the superspace belongs to a trajectory with associated momenta $p_{b} = \nabla^{(c)}_{b} S$ and velocity 
\begin{linenomath}
\begin{equation}\label{eq:Vilenkin-velocity}
    \dot{h}^{a} = 2N \,\nabla^{(c)\, a} S\,,
\end{equation}
\end{linenomath} 
that depends on the choice of $N(t)$ from the foliation. Here, we can infer the form of the time derivative
\begin{linenomath}
\begin{equation}\label{eq:deftimeVilenkin}
    \dtau{} = 2N \nabla^{(c)} S \cdot \nabla^{(c)}\,,
\end{equation}
\end{linenomath}
which will come into play later. The points that satisfy $\nabla^{(c)} S =0$ separate the classically allowed and forbidden regions, breaking down the semiclassical approximation. By requiring that each hypersurface is crossed only once by the congruence of trajectories, {i.e.,} 
\begin{linenomath}
\begin{equation}\label{eq:Vilekin-crossing-hypersurfaces}
    \dot{h}^a\, d\Sigma^{(c)}_a > 0\,, 
\end{equation}
\end{linenomath}
then the probability density
\begin{linenomath}
\begin{equation}
    dP = j^{(c) \, a}\, d\Sigma^{(c)}_a
\end{equation}
\end{linenomath}
is positive semi-definite, thus the Universe wave function can be properly normalized. The same can be implemented for a wave function that is a superposition $\sum_k \Psi_k$ of terms defined as in \eqref{eq:defVilenkinSemiclassical} when the condition \eqref{eq:Vilekin-crossing-hypersurfaces} is satisfied for each $k$, such that the total probability is conserved.

One could then wonder if a similar implementation is possible in the more general case, when only a part of the Universe is semiclassical and the rest must be described in a full quantum picture. Vilenkin examined the case in which the quantum variables (labeled by $q^{\nu}$ with $\nu = 1,...,m$) represent a small quantum subset, with negligible effects on the semiclassical variables ($h^a$ with $a = 1,..,n-m$) dynamics. The full Wheeler--DeWitt equation then becomes
\begin{linenomath}
\begin{equation}\label{eq:vilenkinWDW-quantum}
   \left( -\hbar^2 \nabla^{(c)\,2} +U^{(c)} +\hat{H}^{(q)}\right) \Psi =0 \,,
\end{equation}
\end{linenomath}
where using the previous notation $-\hbar^2 \nabla^{(c)\,2} +U^{(c)} = \hat{H}^{(c)}$ is given neglecting the quantum variables and their conjugate momenta, which instead appear in $\hat{H}^{(q)} = -\hbar^2 \nabla^{(q)\,2} + U^{(q)}$ (here the the superscript $^{(q)}$ refers to the quantum components). At the same time, the semiclassical part is assumed to satisfy its own WDW Equation \eqref{eq:Vil-WDW-semicl}, thus obtaining a system of coupled equations for the two sectors dynamics. This separation is backed both by the hypothesis on the smallness of the quantum subsystem, expressed as
\begin{linenomath}
\begin{equation}\label{eq:VilenkinHqsmall}
    \frac{\hat{H}^{(q)} \Psi}{\hat{H}^{(c)} \Psi} = \ord{\hbar} \,,
\end{equation}
\end{linenomath}
and by the independence between the two sets, namely
\begin{linenomath}
\begin{gather}
    G_{ab} (h,q) = G_{{ab}}(h) +\ord{\hbar}\,,\\
    G_{a\nu} = \ord{\hbar}\,.\label{eq:cond-Gab-mixed}
\end{gather}
\end{linenomath}
In other words, we are assuming $G_{ab}$ to be dependent on the semiclassical variables only, {and the two subspaces to be approximately orthogonal, since any mixed term of the supermetric (being the index $a$ for the semiclassical variables and $\nu$ for the quantum variables) is of higher order in the perturbative expansion}; it follows that higher order terms will not appear inside $\nabla^{(c)\,2}= G_{ab} \nabla^{(c)\,a} \nabla^{(c)\,b}$ {(see for example the applications in Sections}~\ref{sec:appl-vilenkin-bianchiI} and \ref{sec:appl-vilenkin-bianchiIX}).
Following these hypotheses, the wave function can be separated in
\begin{linenomath}
\begin{equation}
    \Psi (h,q) = \psi (h) \chi (q,h) = A(h) e^{\frac{i}{\hbar} S(h)} \chi(h,q).
    \label{eq:ansatz-vilenkin}
\end{equation}
\end{linenomath}
We observe that this ansatz shares similarities with both the WKB and BO approximations. Considering the former, this is due to the presence of a complex exponential with a small parameter that can lead the expansion; instead, the latter is due to a separation of a purely semiclassical sector and the quantum one, as explained in more detail in Section~\ref{sec:WKB}. Actually, Vilenkin's proposal can be reformulated as a special case of a BO-like approximation with expansion in $\hbar$, {i.e.,} the semiclassical expansion, as discussed in Section~\ref{ssec:non-unitarity}. 

We here use again the Planck constant for the expansion, instead of the parameter proportional to $\hbar$, as mentioned before. Since the previous hypotheses hold, the semiclassical function $\psi$ must satisfy Equation~\eqref{eq:Vil-WDW-semicl}, giving Equations~\eqref{eq:Vilenkin-HJ} and \eqref{eq:VilenkinAmplitude} respectively at $\ord{\hbar^0}$ and $\ord{\hbar}$. Meanwhile, the quantum function $\chi$ inherits a different dynamics from Equation~\eqref{eq:vilenkinWDW-quantum}, that is
\begin{linenomath}
\begin{equation}
    -\hbar^2 \nabla^{(c)\, 2} \chi -2\hbar^2 A^{-1} (\nabla^{(c)} A) \cdot \nabla^{(c)}\chi - 2i\hbar\, (\nabla^{(c)} S)\cdot \nabla^{(c)} \chi +\hat{H}^{(q)} \chi =0 \,,
\end{equation}
\end{linenomath}
where we can observe that all terms except the last two are of higher order in the expansion parameter ($H^{(q)}$ is of order $\hbar$ due to assumption \eqref{eq:VilenkinHqsmall}). Thus, at $\ord{\hbar}$, we obtain
\begin{linenomath}
\begin{equation}\label{eq:VilenkinDinamicaChi}
    \hat{H}^{(q)} \chi = 2i\hbar \,(\nabla^{(c)} S)\cdot \nabla^{(c)} \chi \,.
\end{equation}
\end{linenomath}
Multiplying \eqref{eq:VilenkinDinamicaChi} by $N(t)$ and using the same time derivative \eqref{eq:deftimeVilenkin} defined for the semiclassical universe, it becomes
\begin{linenomath}
\begin{equation}
    i\hbar \dtau{\chi} = N \hat{H}^{(q)} \chi \,,
    \label{schro}
\end{equation}
\end{linenomath}
namely a functional Schr\"{o}dinger equation for the matter wave function. 

{It is worth noting that the definition introduced above for $\partial_{\tau}$ is very close to the notion of a composite derivative $\partial_{\tau}\equiv \frac{d h^a}{d\tau} \,\partial_{h^a}$ applied to the quantum wave function. This fact can be easily realized by  recalling that $\partial_{h^a} S_0$ is just the conjugate momentum $p_a$ and, hence, it is enough to write down the first Hamilton equation (obtained variating the classical action with respect to $p_a$) to arrive to the desired statement. By other words, the time dependence of the quantum wave function is recovered in the approach proposed in}  \cite{bib:vilenkin-1989}, {by means of the dependence that the quasi-classical variables $h^a$ acquire, at the leading order, on the label time of the space-time slicing. Clearly, it is also possible and discussed in} \cite{bib:vilenkin-1989} {that one of the $h^a$ themselves is chosen as time coordinate to describe the system evolution., suitably choosing the lapse function $N(t)$.}
{It is also useful to stress that, as we will see later in the considered specific applications, the form of the supermetric $G^{ab}$ as a function of $h^a$ is sensitive to the specific set of adopted configurational variables to describe the studied cosmological model. However, we can observe that any variable among the $h^a$'s, which is related to the Universe volume, acquires a different signature (say a time-like one) different from all the other ones (regarded as space-like coordinates)}  \cite{bib:dewitt1-1967,bib:dewitt2-1967,bib:dewitt3-1967,bib:giovannetti-montani-2022}.
{Independently from the specific form of $G^{ab}$ and $H^{(q)}$, the important point to make safe the model self-consistence is that the semiclassical metric $G^{ab}$ and the quantum one, fixed by the form of $H^{(q)}$ itself, live in orthogonal spaces, i.e. cross terms 
in the supermetric with a classical index and a quantum one must be of higher order in the present formulation, as expressed by} \eqref{eq:cond-Gab-mixed}. 

{Explicit examples of this classical-from-quantum variable separation are given below, see Sections} \ref{sec:appl-vilenkin-bianchiI} and \ref{sec:appl-vilenkin-bianchiIX}. {We consider here both the situations in which this separation takes place between the gravitational degrees of freedom, e.g., Universe volume taken as a quasi-classical variable and space anisotropies as quantum variables, as well as the case in which the same separation concerns quantum matter living on a quasi-classical space-time. This last situation is of particular physical relevance since, as we shall see below, its analysis to the next order of approximation in the order parameter corresponds to the study of quantum gravity corrections to standard quantum field theory.}

Differently from the {purely semiclassical} case, two probability currents now emerge. The one including the semiclassical sector is
\begin{linenomath}
\begin{equation}\label{eq:Vil-semicl-curr}
    j^a = |\chi|^2 |A|^2 \nabla^{(c)\,a} S \equiv j^{(c)\,a}\, \rho_{\chi},
\end{equation}
\end{linenomath}
where $j^{(c)\,a}$ is the same as \eqref{eq:VilenkinCurrent} and $\rho_{\chi} = |\chi|^2$ is the probability distribution of the quantum variables computed on the semiclassical trajectories. 
For the quantum components instead we find
\begin{linenomath}
\begin{equation}\label{eq:Vil-quantum-curr}
    j^{\nu}= -\frac{i}{2} |A|^2 \Bigl(\chi^* \nabla^{(q)\,\nu} \chi - \chi \nabla^{(q)\,\nu} \chi^*\Bigr) = \frac{1}{2} |A|^2 j_{\chi}^{\nu},
\end{equation}
\end{linenomath}
associated to the distribution  $\rho_{\chi}$, where $j_{\chi}^{\nu}$ is a Klein--Gordon-like current. From the conservation of both the total current $\nabla^{(c)}_a j^a + \nabla^{(q)}_{\nu} j^{\nu} = 0$ and the semiclassical current $\nabla^{(c)}_a j^{(c)\,a}=0$, given by the full WDW Equation~\eqref{eq:vilenkinWDW-quantum} and assumption \eqref{eq:Vil-WDW-semicl} respectively, we can state that at the leading order the following equation holds
\begin{linenomath}
\begin{equation}\label{eq:vil-continuity}
    \dtau{\rho_{\chi}} + N \nabla^{(q)}_{\nu} j^{\nu}_{\chi} =0 \,,
\end{equation}
\end{linenomath}
which is a continuity equation for the quantum variables. Moreover, both $\rho_c$ and $\rho_{\chi}$ can be normalized on their respective subspaces by requiring $\int d\Sigma^{(c)} \rho^{(c)} =1$ and $\int d\Omega^{(q)} \rho_{\chi} =1$, being $d\Sigma = d\Sigma^{(c)}\, d\Omega^{(q)}$ the total surface element on the equal-time surfaces identified with the foliation. In this way, the standard probabilistic interpretation is recovered for $\psi$ when such a separation in semiclassical and quantum variables is valid.

However, there is still one case to discuss, that is when in such a framework one (or more) quantum variables become semiclassical at later time. This means that the two subsets change: starting from an initial wave function of the form \eqref{eq:ansatz-vilenkin}, we have $\phi_k \chi_k \rightarrow \sum_l \phi_{k}(h') \chi_{kl}(h',q')$, the new semiclassical set $\{h'\}$ having increased by one variable and the quantum one $\{q'\}$ decreased by one variable. The sum is explained by the transition during which each semiclassical trajectory branches into many trajectories, each one for a different initial condition of the \lq\lq new semiclassical'' variable. For this reason, one has to impose a unitarity (normalization) condition on the semiclassical current $j^{(c)\,a}$
\begin{linenomath}
\begin{equation}
    \int d\Sigma_{k\,a}^{(c)}\, j_k^{(c)\,a} = \sum_{l} \int d\Sigma_{kl \,a}^{(c)} \, j^{(c)\,a}_{kl},
\end{equation}
\end{linenomath}
that is satisfied only at an approximate level, {i.e.,} when the cross terms can be neglected. It should be stressed that the division itself between the two subspaces is heavily dependent on the considered case and almost arbitrary in a certain footing, leading to an approximate concept of unitarity for the Universe.

\subsection{Boundary Conditions for the Cosmological Wave Function}\label{ssec:boundary}
Vilenkin's work provides a meaningful description at the typical scale of the quantum subsystem of the Universe. One related point concerns how to impose boundary conditions on the wave function \eqref{eq:ansatz-vilenkin}, which has led to ample discussion in the literature. Vilenkin himself had previously studied this issue \cite{bib:vilenkin-1982,bib:vilenkin-1983,bib:vilenkin-1986}, developing the so-called \emph{tunneling proposal}: he constructed a wave function describing an ensemble of Universes that tunnel from \lq\lq nothing'' to a de Sitter space by implementing a similar expansion of $\Psi$ \eqref{eq:ansatz-vilenkin} and choosing the purely expanding solution.

A different implementation is the one by Hartle--Hawking \cite{bib:hartle-hawking-1983}, also known as the \emph{no-boundary} proposal. The wave function for a closed Universe is constructed in the Euclidean path integral approach by integrating over all the possible compact 4-geometries corresponding to a certain induced metric $h_{ij}$ on a spacelike boundary {(see also discussion in} \cite{bib:page-2007}); the resulting wave function can be shown to approximately satisfy the WDW equation, {whose} {corresponding} Hamiltonian is required to be a Hermitian operator. 

In this respect, the path integral approach \cite{bib:feynman-path} represents an alternative formulation of gravity as a quantum field theory and it has been widely discussed in relation to the problems of time and unitarity \cite{bib:barvinsky-1990,bib:barvinsky-1993,bib:vilenkin-1994-rassegna,bib:amaral-bojowald-2016}. We mention that, actually, Vilenkin's tunneling proposal can be reformulated in the Lorentzian path integral formalism \cite{bib:vilenkin-1984}. {The WKB implementation also allows to study the probability of tunneling from a false vacuum to a true vacuum state from the Wheeler--deWitt equation (see} \cite{bib:kristiano-2019} {and references within).} We will here focus on the Dirac quantization method only, however an interesting discussion between the two schemes can be found in \cite{bib:halliwell-1988} where, using the Lorentzian path integral, the WDW equation is uniquely recovered in the minisuperspace via a particular gauge fixing on the values of $h_{ij}$ and $N$, that solves the operator-ordering ambiguity of the Dirac scheme. Some sort of WKB procedure \emph{a la} Vilenkin can also be included in the path integral formalism to study the boundary conditions, see for example \cite{bib:lehners-2015,bib:bramberger-2017,bib:jonas-lehners-2021} (Lorentzian), finding in some cases different features with respect to the Hartle--Hawking interpretation. 

\section{Validation of the Vilenkin Proposal for the Bianchi I Cosmology}
\label{sec:appl-vilenkin-bianchiI}

One of the most interesting open questions in theoretical cosmology concerns how a primordial quantum universe reaches a classical isotropic limit \cite{bib:montani-primordialcosmology}. The reason to hypothesise a very general morphology of the universe near the singularity (for a big bounce picture of the Bianchi I model see \cite{bib:ashtekar-2009,bib:cianfrani-marchini-montani-2012,bib:moriconi-montani-2017,bib:montani-marchi-moriconi-2018,bib:montani-schiattarella-2021}) relies on the request to address the quantum cosmological problem within the Bianchi homogeneous framework \cite{bib:cianfrani-canonicalQuantumGravity}. These models are characterized by the preservation of the space-line element under a specific group of symmetry, and are collected in the so-called Bianchi classification. 

\subsection{The Minisuperspace Dynamics of Bianchi Universes}\label{ssec:minisuperspace-bianchi}

\begin{figure}[H]
  \includegraphics[width=9cm]{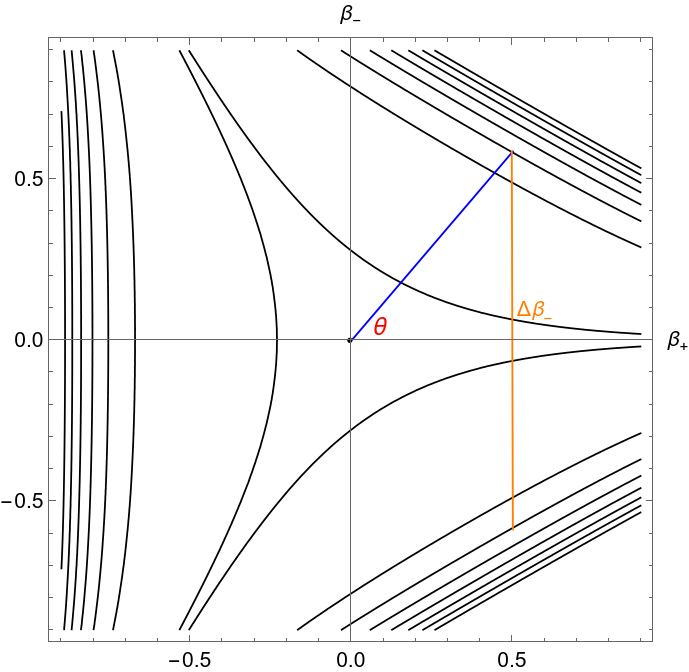}
  \caption{Description of Bianchi IX potential isocurve on which is marked the corner structure. Here, $\theta$ describes the width of the $\beta_+$ channel.}
  \label{triangularfig}
\end{figure}

The most general homogeneous model is the Bianchi IX model \cite{bib:belinsky-1970,bib:landau2,bib:montani-primordialcosmology}, also called Mixmaster model \cite{bib:misner-1969-mixmaster} {(for a recent semiclassical discussion see} \cite{bib:brizuela-uria-2022}), that has a relevant role in the study of the cosmological dynamics. Despite its spatial homogeneity, it presents typical features of the generic cosmological solution such as a chaotic time evolution of the cosmic scale factors near the singularity \cite{bib:imponente-montani-2001}. This corresponds to an infinite sequence of bounces of the point particle, in the Hamiltonian representation, against the time-dependent potential walls which can be shown to induce an ergodic evolution in the Misner--Chitre variables. The standard dynamics in the central region of the potential well are then restored~\cite{bib:montani-primordialcosmology} once it escapes the small oscillations configuration. However, in the asymptotic limit to the cosmological singularity, the potential term of Bianchi IX dynamics has the morphology of an equilateral triangle and three open corners appear in the vertices, which correspond to the non-singular Taub cosmology \cite{bib:misner-taub-1969}, see Figure~\ref{triangularfig} \cite{bib:montani-chiovoloni-2021,bib:deangelis-montani-2022-emergence}.

This kind of cosmology defines the limit of Bianchi IX dynamics when two scale factors are considered equal over the three possible independent ones.
The importance of the Hamiltonian formulation of the Mixmaster model (see \cite{bib:misner-1969-mixmaster}) using the ADM description, relies on the fact that it is possible to reduce the dynamics to the two-dimensional point particle. 
We start with the line element of the model in the Misner picture
\begin{linenomath}
\begin{equation}
    ds^2= N(t)^2dt^2-\eta_{ab}\omega^a \omega^b,
    \label{linelementbianchi}
\end{equation}
\end{linenomath}
where $\omega^a=\omega^a_{\alpha}dx^{\alpha}$ is a set of the three invariant differential forms that fixes the geometry of the considered Bianchi model, $N(t)$ is the lapse function and $\eta_{ab}$ is defined as \mbox{$\eta_{ab}=e^{2\alpha}(e^{2\beta})_{ab}$}. The choice of these variables allows us to separate the isotropic contribution expressing the volume of the universe related to $\alpha$, {i.e.,} for $\alpha \rightarrow -\infty$ the initial singularity is reached, from the gravitational degrees of freedom $\beta_+$, $\beta_-$ contained in the matrix $\beta_{ab}=diag(\beta_+ +\sqrt{3}\beta_-, \beta_+-\sqrt{3}\beta_-, -2\beta_+)$ acting as the anisotropies of this model. Moreover, the introduction of the Misner variables makes the kinetic term in the Hamiltonian diagonal. We can rewrite the superHamiltonian constraint as
\begin{linenomath}
\begin{equation}
    H_{IX}=\frac{\kappa}{3(8\pi)^2}e^{-3\alpha}(-p_{\alpha}^2+p_+^2 +p_{-}^{2}\color{black} +\mathcal{V}+\Lambda e^{6 \alpha})=0,
    \label{H9}
\end{equation}
\end{linenomath}
where $\kappa=8\pi G/c^4$ is the Einstein constant and the potential $\mathcal{V}$ takes the form
\begin{linenomath}
\begin{equation}
    \mathcal{V}\equiv -\frac{6(4 \pi)^4}{\kappa^2}\eta R^{(3)} = \frac{3(4\pi)^4}{\kappa^2}e^{4\alpha}V_{IX}(\beta_{\pm}),
\end{equation}
\end{linenomath}
where the spatial scalar of curvature generates the Bianchi IX potential term depending only on the anisotropies
\begin{linenomath}
\begin{equation}
    V_{IX}(\beta_{\pm})=e^{-8\beta_+}-4e^{-2\beta_+}\cosh({2\sqrt{3}\beta_-})+2e^{4\beta_+}[\cosh({4\sqrt{3}\beta_-})-1].
\end{equation}
\end{linenomath}
This function has the symmetry of an equilateral triangle with steep exponential walls and three open angles. The expressions for the equipotential lines for large values of $|\beta_+|$ and small $|\beta_-|$ are
\begin{linenomath}
\begin{equation}
   V_{IX}(\beta_{\pm})\sim 
   \begin{cases}
   e^{-8\beta_+}  & \beta_+\rightarrow -\infty, \; |\beta_-|\ll 1 \\
   48e^{4\beta_+} \beta_-^2  & \beta_+\rightarrow +\infty, \; |\beta_-|\ll 1
   \end{cases}
\end{equation}
\end{linenomath}
while close to the origin, for $\beta_{\pm}\rightarrow 0$,
\begin{linenomath}
\begin{equation}
   V_{IX}(\beta_{\pm})\sim \beta_+^2+\beta_-^2.
\end{equation}
\end{linenomath}
The Hamiltonian approach provides the following equations of motion
\begin{linenomath}
\begin{equation}
    \dot{\alpha}=N\frac{\partial H_{IX}}{\partial \alpha}, \qquad \dot{p}_{\alpha}=N\frac{\partial H_{IX}}{\partial \alpha},
    \label{eqmot1}
\end{equation}
\end{linenomath}
\begin{linenomath}
\begin{equation}
    \dot{\beta}_{\pm}=N\frac{\partial H_{IX}}{\partial p_{\pm}}, \qquad \dot{p}_{\pm}=N\frac{\partial H_{IX}}{\partial \beta_{\pm}}.
    \label{eqmot2}
\end{equation}
\end{linenomath}
One recognizes that the dynamics of the universe towards the singularity is mapped into the motion of a particle that lives on a plane inside a closed domain and bounces against the potential wall.

The canonical quantization of the system consists of the commutation relations
\begin{linenomath}
\begin{equation}
    [\hat{q}_a,\hat{p}_b]=i\delta_{ab},
\end{equation}
\end{linenomath}
which are satisfied for $\hat{p}_a=-i\frac{\partial}{\partial q_a}= -i\partial_a$ where $(a,b = \alpha, \beta_+, \beta_-)$ adopting natural units. By replacing the canonical variables with the corresponding operators, the quantum behaviour of the universe is given by the quantum version of the superHamiltonian constraint \eqref{H9}, {i.e.,} the WDW equation for the Bianchi IX model
\begin{linenomath}
\begin{equation}
	\hat{H}_{IX}\Psi(\alpha, \beta_{\pm})=\bigg[\partial_{\alpha}^2-\partial_{+}^2-\partial_-^2+\frac{3(4\pi)^4}{\kappa^2}e^{4\alpha}V_{IX}(\beta_{\pm})\bigg]\Psi(\alpha,\beta_{\pm})=0,
	\label{Hhat}
\end{equation}
\end{linenomath}
where $\Psi(\alpha,\beta_{\pm})$ is the wave function of the universe providing information about its physical state.
Therefore, following the step in \cite{bib:vilenkin-1989} we can obtain the probability distribution for the wave function of the universe, that reads as
\begin{linenomath}
\begin{equation}
    \rho(\alpha, \beta_{\pm},t)=\rho^{(c)}(\alpha,t)\,\rho_{\chi}(\alpha,\beta_{\pm}(t),t),
    \label{prob}
\end{equation}
\end{linenomath}
where in particular $\rho^{(c)}(\alpha, t)=|A(\alpha(t))|^2$ is related to the components of the classical space and $\rho_{\chi}(\alpha, \beta_{\pm}(t),t)=|\chi(\alpha,\beta(t),t)|^2$ to those in the quantum subspace, as explained in Section~\ref{sec:vilenkin}.

In the subsection below, we will focus on the Bianchi I model in which the structure constants and so the spatial curvature $R^{(3)}$ vanishes. Hence, the associated superHamiltonian constraint in vacuum read as
\begin{linenomath}
\begin{equation}\label{eq:H-bianchi-I}
H_{I}=\frac{\kappa}{3(8\pi)^2}e^{-3\alpha}(-p_{\alpha}^2+p_+^2+p_-^2)=0.
\end{equation}
\end{linenomath}
This cosmology is the natural extension of the FRLW model with $k=0$ generalizing an homogeneous flat Universe.

\subsection{Implementation of the WKB Approach in the Minisuperspace of Bianchi I}\label{ssec:conto-bianchi-I}

Let us consider the case of the Bianchi I model, {i.e.,} with WDW expressed by the constraint \eqref{eq:H-bianchi-I}. As mentioned in Section~\ref{ssec:problemtime}, the time definition for the model can be implemented before or after quantization, namely the RPSQ and Vilenkin's proposal, which present striking differences.
The Vilenkin proposal (Section~\ref{sec:vilenkin}) is more feasible, namely it avoids the square root non-local Hamiltonian operator emerging from the former but the probabilistic interpretation is achieved only after performing the semiclassical limit and, in this sense, it can not be seen as a fundamental approach. However, the role of the time-like variables itself can make the two schemes comparable. In order to determine under which restrictions the Vilenkin representation of the Universe volume dynamics becomes predictive, it has been shown in \cite{bib:agostini-cianfrani-montani-2017} a rigorous comparison of the two quantization methods carrying out the probabilistic interpretation of the wave function for the Bianchi I cosmology in which $R^{(3)}=0$. 

We recall that Vilenkin suggested a semiclassical approximation of the wave function to achieve a proper probabilistic interpretation due to the emergence of time. This does not happen for the definition of a scalar product from a conserved current suggested by DeWitt. Hence, considering a Bianchi I model in the presence of a matter contribution and achieving the Schr\"odinger equation describing the motion of a free particle in the $(\beta_+,\beta_-)$ plane, we arrive at the following result
\begin{linenomath}
\begin{equation}
    \Psi(\alpha, \beta_a)=\frac{e^{-\frac{i}{\hbar} \int_{\alpha_0}^{\alpha}d\alpha'\sqrt{\mu^2(\alpha)}}}{^4\sqrt{\mu^2(\alpha)}}\int_{\mathbb{R}^2}\frac{d^2p}{2\pi \hbar}e^{-\frac{i}{2 \hbar}(p_+^2+p_-^2)\int_{\alpha_0}^{\alpha}d\alpha'\frac{1}{\sqrt{\mu^2(\alpha)}}} \cdot e^{\frac{i}{\hbar}p_a \beta_a}\tilde{\chi}(\alpha_0,p_a),
    \label{vilenkinphi}
\end{equation}
\end{linenomath}
where the subscript $_a$ stands for $(+,-)$ and $\tilde{\chi}(\alpha_0,p_a)$ determines initial conditions. The matter contribution is encoded in the term $\mu^2$ as
\begin{linenomath}
\begin{equation}
    \mu^2(\alpha)=\sum_w \mu^2_{w}\, e^{3(1-w)\alpha},
\end{equation}
\end{linenomath}
where the sum contains all the fluids components characterized by different values of $w$, while $\mu^2_w$ are constants.
It is important to stress that with the BO approximation we are assuming $\alpha$ as the slow variable whereas $\beta_+$ and $\beta_-$ are the fast ones. The validity of both Vilenkin's semiclassical expansion and BO approximation, which will be discussed in detail in Sections~\ref{sec:WKB} and \ref{ssec:BO}, implies that we admit a decomposition of the wave function as  
\begin{linenomath}
\begin{equation}
\Psi(\alpha,\beta_a)=\exp{ \left(\frac{i}{\hbar}\sum_{n=0}(\hbar)^n S_{n} \right)}
\end{equation}
\end{linenomath}
and the following conditions hold
\begin{linenomath}
\begin{align}
    \biggl{|}\frac{1}{\mu^3(\alpha)}\frac{d\mu^3(\alpha)}{d\alpha}&\biggl{|} \ll \frac{4}{\hbar},\\
    \hbar |S_2(\alpha)| \ll |S_1(\alpha)| \qquad &\text{and} \qquad \hbar |S_2(\alpha)|\ll 1.
\end{align}
\end{linenomath}
Moreover, the integral over the momentum space extends over those values for which
\begin{linenomath}
\begin{align}
    (p_+^2+p_-^2) &\neq 0,\\
    (p_+^2+p_-^2) &\ll \mu^2(\alpha).
    \label{secondcondition}
\end{align}
\end{linenomath}
Now, the BO approximation implies that near a value for which \eqref{secondcondition} holds, we need a wave packet for the initial conditions sufficiently peaked, for simplicity a Gaussian distribution of the form
\begin{linenomath}
\begin{equation}
    \tilde{\chi}(\alpha_0,p_a)=\frac{1}{\sqrt{\pi \sigma_+\sigma_-}}e^{-\frac{(p_+-\bar{p}_+)^2}{2\sigma_+^2}}e^{-\frac{(p_--\bar{p}_-)^2}{2\sigma_-^2}},
\end{equation}
\end{linenomath}
The aim is to check whether the functional form of the wave functions obtained from the two formalisms, or their associated probabilities, coincide. In order to do this, we need the Klein--Gordon-like time-independent inner product, achieved from the RPSQ approach.

Following the steps described in \cite{bib:cianfrani-canonicalQuantumGravity,bib:thiemann-2006-rpsq} the resulting Schr\"odinger equation is
\begin{linenomath}
\begin{equation}
    i\hbar \frac{\partial}{\partial \alpha}\Phi(\beta_a,\alpha)=\sqrt{-\hbar^2\left(\frac{\partial^2}{\partial \beta_+^2}+\frac{\partial^2}{\partial \beta_-^2}\right)+\mu^2(\alpha)} \; \Phi(\beta_a, \alpha),
\end{equation}
\end{linenomath}
in which $\lim_{\alpha \rightarrow-\infty}\mu^2(\alpha)=\mu^2_1$ and we denoted $\Phi$ as the wave function of the RPSQ. Hence, via inverse Fourier transform a generic solution can be formally found as
\begin{linenomath}
\begin{equation}
    \Phi(\beta_a,\alpha)=e^{-\frac{i}{\hbar}\int_{\alpha_0}^\alpha d\alpha ' \sqrt{-\hbar^2 \Delta_{\pm}+\mu^2(\alpha)}}\Phi(\beta_a,\alpha_0),
    \label{phi}
\end{equation}
\end{linenomath}
where $|p_a|^2=-\hbar^2 \Delta_{\pm} = -\hbar^2 \left(\frac{\partial^2}{\partial \beta_+^2}+\frac{\partial^2}{\partial \beta_-^2}\right)$. Now, to compare the two formulations, we need to identify the same time variable. In particular, we need the two lapse functions (one from RPSQ and the other one from Vilenkin's proposal) to be the same
\begin{linenomath}
\begin{equation}
    \frac{3 c \mathcal{K}}{4\pi GT}\frac{e^{3\tau}}{\sqrt{p_+^2+p_-^2+\mu^2(\tau)}}=\frac{3c \mathcal{K}}{4 \pi G T}\frac{e^{3\tau}}{\sqrt{\mu^2(\tau)}},
    \label{equalN}
\end{equation}
\end{linenomath}
where $\mathcal{K}=\int d^3x |\text{det}(e_i^{(a)}(x^k))|$, the vectors $e_i^{(a)}$ constitute the so-called \emph{frame} and $\alpha=t/T=\tau$ in which the constant $T$ can be defined in terms of fundamental constants, {e.g.,} it can be chosen proportional to the Planck length. The above equation is effectively valid if $p_+^2+p_-^2 \ll \mu^2(\tau)$. An issue arises if we promote $\beta_a$ to quantum operators since the lapse function $N_{RPSQ}$ (on the left-hand side) becomes an operator acting on the wave function. For this reason, we need to replace it by its expectation value. However, in Bianchi I, $p_a$ are essentially constants of motion and we can treat them as numbers. Now, we are able to choose a range of $\tau$ such that the semiclassical approximation is valid, namely $\tau_S$. Hence, by normalizing \eqref{phi} with respect to the inner product near the singularity, such that $\mu^2 \rightarrow \mu^2_1$ becomes time-independent \cite{bib:mostafazadeh-2002}, implementing it with the BO approximation, we can~write
\begin{adjustwidth}{-\extralength}{0cm}
\begin{equation}
    \Phi(\beta_a, \tau)\approx \frac{e^{-\frac{i}{\hbar}\int_{\tau_S}^{\tau} d\tau ' \sqrt{\mu^2(\tau ')}}}{^4\sqrt{\mu^2_1}}\int_{\mathbb{R}^2}\frac{d^2p}{(2\pi\hbar)\sqrt{2\pi \sigma_+ \sigma_-}}\frac{e^{-\frac{i}{2\hbar}(p_+^2+p_-^2)\int_{\tau_S}^{\tau}d\tau ' \frac{1}{\mu^2(\tau ')}}}{^4\sqrt{1+\frac{p_+^2+p_-^2}{\mu^2_1}}}e^{\frac{i}{\hbar}p_a  \beta_a}e^{-\frac{(p_+-\bar{p}_+)^2}{2\sigma_+^2}}e^{-\frac{(p_--\bar{p}_-)^2}{2\sigma_-^2}}.
    \label{RPSQphi}
\end{equation}
\end{adjustwidth}

Two main differences are noticed comparing \eqref{RPSQphi} with \eqref{vilenkinphi}. In Equation~\eqref{vilenkinphi}, the factor $\left(1+\frac{p_+^2+p_-^2}{\mu^2_1}\right)^{-1/4}$ is not present and $(\mu^2_1)^{-1/4}$ is replaced by $(\mu^2(\tau))^{-1/4}$. However, we achieve the same probability of finding the Universe in a region of the plane $(\beta_+, \beta_-)$ for both approaches, if the spectra of the corresponding momenta span sufficiently small values. In this way, the contribution of the anisotropies to the total energy is negligible with respect to the matter part. In other words, for the Vilenkin approach we need to impose a constraint on the anisotropies variables phase space, namely that they exhibit a \lq \lq light~dynamics''.

\section{Implementation of the Vilenkin Approach to the Bianchi IX \lq \lq Corner''}\label{sec:appl-vilenkin-bianchiIX}

Let us now analyse the well-known Bianchi IX \lq\lq corner'' configuration \cite{bib:montani-primordialcosmology} implementing the WKB idea to separate the quasi-classical component from the \lq \lq small'' variable $\beta_-$.
Thus, to describe Bianchi IX's dynamics near the singularity using the Misner variables and the Vilenkin approach, we consider as an initial condition for the point-universe the right corner of the potential $V_{IX} \sim 48e^{4\beta_+}\beta_-^2$, where $\beta_+ \rightarrow +\infty$ and $|\beta_-| \ll 1$, therefore, $\alpha$ and $\beta_+$ have to be semiclassical variables while $\beta_-$ quantum. In the following analysis (see~\cite{bib:montani-chiovoloni-cascioli-2020}) we will include a massless scalar field $\phi$ for which $\dot{\phi}\ll U(\phi)$, and we will assume a synchronous frame $N(t)=1$. 

Substituting the ansatz \eqref{eq:ansatz-vilenkin} and using the conditions above in the WDW equation, Equation~\eqref{eq:VilenkinAmplitude} becomes
\begin{linenomath}
\begin{equation}
    2\left(\partial_{\alpha}A\, \partial_{\alpha}S-\partial_{+}A\, \partial_+S-\partial_{\phi} A \,\partial_{\phi}S\right)+A\left(\partial^2_{\alpha}S-\partial^2_{+}S-\partial^2_{\phi}S\right)=0,
\end{equation}
\end{linenomath}
associated to the probability density, while the dynamics of a harmonic oscillator with time-dependent frequency and unitary mass reads as
\begin{linenomath}
\begin{equation}
i\hbar \frac{\partial \chi}{\partial \tau}= (\partial_-^2+16e^{4(\alpha+\beta_+)}\beta_-^2)\chi,
\label{schro1}
\end{equation}
\end{linenomath}
if we impose $\omega^2(\tau)\equiv 16e^{4(\alpha+\beta_+)}$ and $\tau=c\int e^{-3\alpha}dt$. Note that in what follows time will be rescaled by a factor 2 as in \cite{bib:montani-chiovoloni-cascioli-2020}. To solve \eqref{schro1} we make use of the invariant method developed in \cite{bib:lewis-1967}. The general solution is given by
\begin{linenomath}
\begin{equation}
    \chi=\sum_n c_n e^{i\alpha_n(\tau)}\phi_n(\beta_-,\tau)=\sum_n c_n\chi_n(\beta_-,\tau),
    \label{chigenerale}
\end{equation}
\end{linenomath}
where $c_n$ are numerical coefficients that weight the different $\chi_n$\vspace{6pt}
\begin{linenomath}
\begin{equation}
c_n=\int d\beta_-\chi_{n}(\beta_-,\tau)\chi_0(\beta_-,\tau),
\end{equation}
\end{linenomath}
\begin{linenomath}
\begin{equation}
    \chi_n(\beta_-,\tau)=\frac{e^{i\alpha_n(\tau)}}{\sqrt{\sqrt{\pi}n!2^n\rho}}h_n\left(\frac{\beta_-}{\rho}\right) e^{\frac{i}{2\hbar}(\frac{\dot{\rho}}{\rho}+\frac{i}{\rho^2})\beta_-^2},
\end{equation}
\end{linenomath}
where the index $_0$ states the initial condition, $h_n$ are Hermite polynomials, $\rho$ satisfies the auxiliary equation
\begin{linenomath}
\begin{equation}
\ddot{\rho}+\omega^2\rho-\rho^{-3}=0,
\label{auxiliary}
\end{equation}
\end{linenomath}
and
\begin{linenomath}
\begin{equation}
    \alpha_n(\tau)=-\left(n+\frac{1}{2}\right)\int^{\tau}_0\frac{1}{\rho^2}d\tau'.
\end{equation}
\end{linenomath}
It is usually complicated to analytically solve \eqref{auxiliary}, but in \cite{bib:lewis-1967} the author developed a method that allows us to have the explicit expression for $\rho$, linear combination of functions $h(\tau)$ and $r(\tau)$ dependent on the considered model
\begin{linenomath}
\begin{equation}
\rho=(\mathcal{W})^{-1}(A^2r^2+B^2h^2+2(A^2B^2-(\mathcal{W})^2)^{\frac{1}{2}}hr)^{\frac{1}{2}},
\end{equation}
\end{linenomath}
where $A^2$, $B^2$ are arbitrary real constants, and $\mathcal{W}$ is the Wronskian.

As a first step, the dynamical evolution of the Mixmaster model could be studied in vacuum, namely the simplest case. {For further studies of Bianchi IX considering a vector field, see} \cite{bib:berkowitz-2021(2), bib:benini-kirillov-montani-2007}.

\subsection{Bianchi IX in Vacuum}\label{ssec:bianchi-IX-vacuum}

Starting from \eqref{schro1} and using \eqref{Hhat}, \eqref{eqmot1}, \eqref{eqmot2} in particular we find
\begin{linenomath}
\begin{equation}
    \alpha(\tau)=\frac{1}{3}\log \left(6|p_{\alpha}|K\right) +2|p_{\alpha}|\tau,
\end{equation}
\end{linenomath}
where $K=\kappa /3(8\pi)^2$.
It is worth noting that, in the calculation above, we adopted the absolute value of $p_{\alpha}$ due to its relation to $\dot{\alpha}$. In fact the expression for $\dot{\alpha}$ (with $\dot{}$ referring to the synchronous time $t$), 
\begin{linenomath}
\begin{equation}
    \dot{\alpha}(t)=-2K p_{\alpha} e^{-3\alpha},
\end{equation}
\end{linenomath}
denotes how much the volume of the universe changes with the synchronous time and it has the opposite sign of $p_{\alpha}$, so that an \emph{expanding} universe is described by $p_{\alpha}<0$. Considering the variable $\tau$, for $0<t<+\infty$ we have $-\infty<\tau<+\infty$.
At the same time, for an expanding universe, the semi-classical variable $\beta_+$ increases toward larger values that means $\dot{\beta}_+(t)>0$ and this, again, translates in $p_+>0$. Therefore, the equation for $\beta_+(\tau)$ is
\begin{linenomath}
\begin{equation}
    \beta_+(\tau)=\beta_0+2|p_{\alpha}|\tau.
\end{equation}
\end{linenomath}
Hence, the frequency for the harmonic oscillator becomes $\omega^2(\tau) \sim C \,e^{m \tau}$, with $m$ and $C$ constants. Now, we can compute the expression for $\rho$ that reads as
\begin{adjustwidth}{-\extralength}{0cm}
\begin{equation}
    \rho=\frac{1}{2m}\sqrt{\pi^2 J_0^2\left(\frac{2\sqrt{C}\sqrt{e^{m\tau}}}{m}\right)+ 64m^2N_0^2\left(\frac{2\sqrt{C}\sqrt{e^{m\tau}}}{m}\right)+8\pi \sqrt{3}m J_0\left(\frac{2\sqrt{C}\sqrt{e^{m\tau}}}{m}\right)N_0\left(\frac{2\sqrt{C}\sqrt{e^{m\tau}}}{m}\right)},
\end{equation}
\end{adjustwidth}
where $J_0$ and $N_0$ represent the Bessel functions of the first and the second kind.

\begin{figure}[H]
  \includegraphics[width=10.5cm]{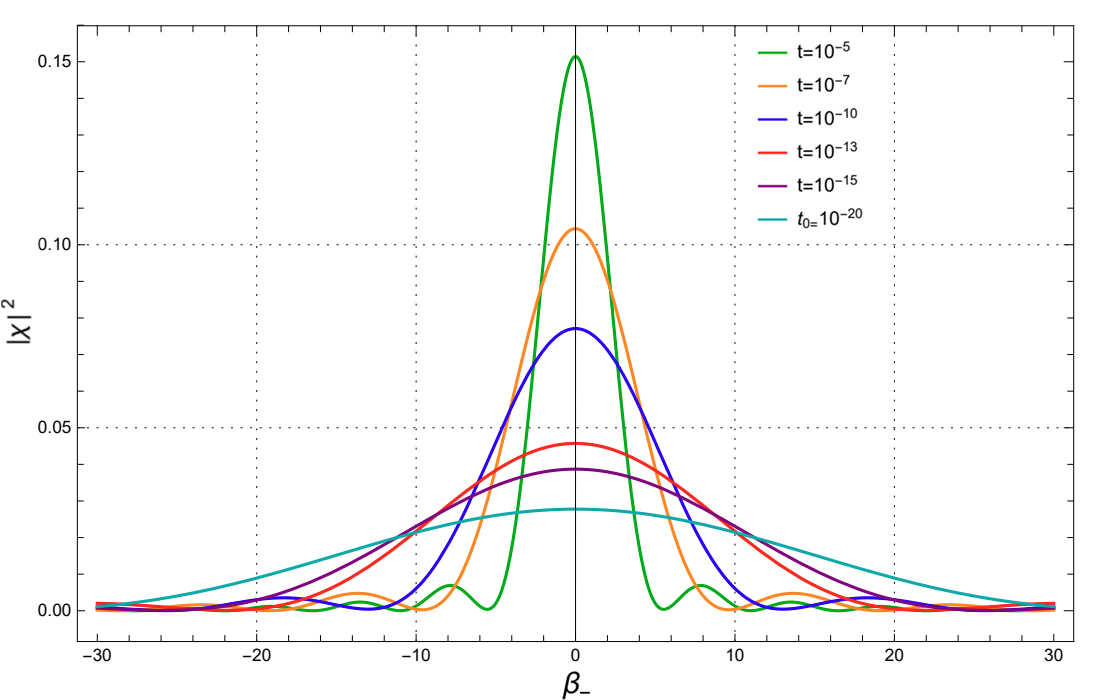}
  \caption{Time evolution of the probability density of the quantum subsystem considering Bianchi IX in the vacuum case with $\beta_+ = 2|p_{\alpha}|\tau$, for an expanding universe. Figure re-elaborated {from} \cite{bib:montani-chiovoloni-cascioli-2020}.}
  \label{vacuumfig}
\end{figure}

To conclude the study of the probability density, firstly we need to compute \eqref{prob} using~\eqref{chigenerale}. We choose $|\chi_0|^2$ such that it has a Gaussian shape peaked around $\beta_-=0$. Figure \ref{vacuumfig} shows the probability density function for different values of the synchronous time variable as a function of the quantum anisotropic variable $\beta_-$.  We observe that, when the point-universe enters the corner, there is a suppression of the quantum variable $\beta_-$, as its standard deviation decays in time. In other words, the Gaussian packet tends to peak around the value $\beta_-=0$. The corner becomes an attractor for the global system dynamics and the point-universe cannot escape anymore. This is the reason why the universe approaches on a good level the Taub model.

The vacuum case can be analysed also for a \emph{collapsing} behaviour of the universe. The dynamical evolution is represented by a decreasing $\beta_+$ for $t \rightarrow +\infty$. 
In this case, the initial assumption is that $\dot{\beta}_+<0$, which translates in $p_+<0$. Therefore, following the same steps, we achieve
\begin{linenomath}
\begin{equation}
    \rho(\tau)= \frac{1}{\sqrt{\omega(\tau)}} =\frac{e^{-\beta_0}}{2},
\end{equation}
\end{linenomath}
in which $\omega(\tau)=4e^{2\beta_0}$ is constant.
The eigenfunctions $\chi_n$ which depend on time through $\rho(\tau)$ are now constant; hence the probability density distribution $|\chi|^2$ is defined simply by choosing its shape at the initial time. This means that it remains constant as the point-universe moves towards the time singularity, namely the point universe goes deeply inside the corner $(\dot{\beta}_+<0)$. Here, the backward evolution of the universe would correspond to a Taub universe, which is no longer a singular cosmology in the past, endowed with a small fluctuating anisotropic degree of freedom in addition to the macroscopic classical universe. Hence, the singular behaviour of the Bianchi IX universe can be removed. This result could have a deep implication, under cosmological hypotheses, on the notion of the cosmological singularity as a general property of the Einstein's equations (see Section~\ref{ssec:bianchi-IX-ihnom} {and for the possible removal of the singularity in loop quantum cosmology (LQC)} see \cite{bib:ashtekar-singh-2006, bib:haro-2012}).

\subsection{Bianchi IX in the Presence of the Cosmological Constant and a Massless Scalar Field}\label{ssec:bianchi-IX-cosmconst}

\begin{figure}[H]
  \includegraphics[width=10.5cm]{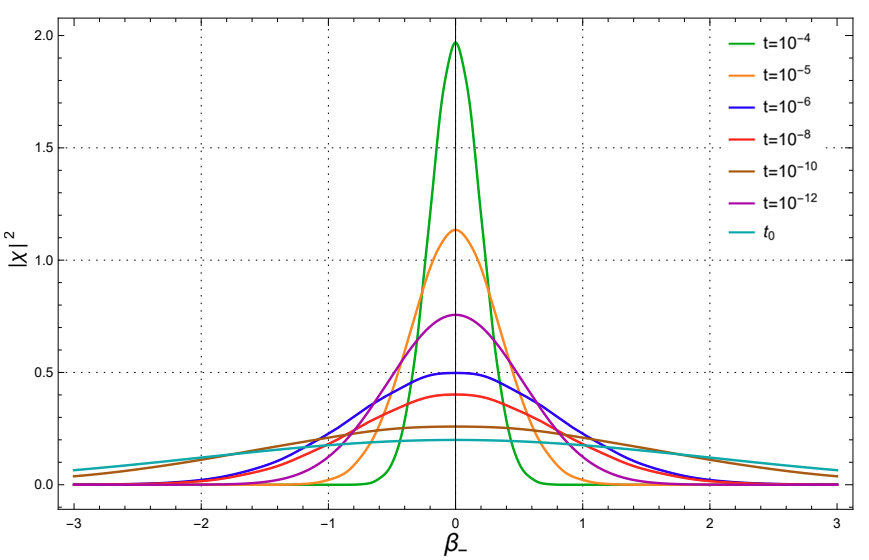}
  \caption{Time evolution of the probability density of the quantum subsystem considering a scalar field and a cosmological constant, in the case of an expanding Universe. Figure re-elaborated {from}~\cite{bib:montani-chiovoloni-cascioli-2020}.}
  \label{notvacuumfig}
\end{figure}

The aim of this analysis is to mimic the behaviour of the Bianchi IX universe if the de Sitter phase (which is associated to the introduction of the cosmological constant $\Lambda$ and the scalar field $\phi$) takes place when the corner evolution is performed by the point-universe. The quantum part of the superHamiltonian $H_q$ does not change with respect to the previous one but we have extra terms in the classical part, namely
\begin{linenomath}
\begin{equation}
    H_0=e^{-3\alpha}K(-p_{\alpha}^2+p_+^2+p_{\phi}^2+\Lambda e^{6\alpha}),
\end{equation}
\end{linenomath}
where $K=\kappa /3(8\pi)^2$. Now, following the same steps of the previous Section \ref{ssec:bianchi-IX-vacuum}, expressions for $\tau(t)$, $\alpha(\tau)$ and $\beta_+(\tau)$ are 
\begin{linenomath}
\begin{equation}
    \tau(t)=\frac{1}{6\sqrt{p_+^2+p_{\phi}^2}}\log\left[\tanh \left(\frac{1}{2}(6K\sqrt{\Lambda}t+J)\right)\right], \qquad \qquad \qquad \,\,\, -\infty<\tau<0
\end{equation}
\end{linenomath}
\begin{linenomath}
\begin{equation}
    \alpha(\tau)=\frac{1}{3}\log \left[\frac{\sqrt{p_+^2+p_{\phi}^2}}{\sqrt{\Lambda}}\sinh \left(2\arctanh\left(e^{6\tau\sqrt{p_+^2+p_{\phi}^2}}\right)\right)\right], \qquad \; -\infty<\alpha<\infty
\end{equation}
\end{linenomath}
\begin{linenomath}
\begin{equation}
\beta_+(\tau)=\beta_0+p_+\tau, \qquad \qquad \qquad \qquad \qquad \qquad \qquad \qquad \qquad \qquad \; \;-\infty<\beta_+<\beta_0.
\end{equation}
\end{linenomath}
This time, the evolution of the probability density function is computed numerically for different values of $t$, as shown in Figure~\ref{notvacuumfig}.

We can conclude that as the universe evolves in time, the variable $\beta_+$ is suppressed while the fully quantum one $\beta_-$ is characterized by a decaying standard deviation. Hence, in the proposed scheme, the universe naturally isotropizes. In other words, starting with a Gaussian shape, its evolution is then approaching a Dirac $\delta$-function around the zero value of $\beta_-$.
Thus, this result offers a new paradigm for the Bianchi IX cosmology isotropization based on the idea that the de-Sitter phase is associated with the corner regime of the model. 

\subsection{Taub Model}\label{ssec:taub}

Another interesting application \cite{bib:deangelis-2020} could be the case of the Taub model \cite{bib:taub-1951} (for a full quantization see \cite{bib:battisti-lecian-montani-2009-taub,bib:catren-2000-taub,bib:berkowitz-2021,bib:berkowitz-2020-taub}) that is the natural intermediate step between Friedmann-Lema\^itre-Robertson-Walker (FLRW), which is invariant under rotations around any axis, and the Bianchi IX universe in which the rotational invariance is absent due to the presence of three different scale factors. Therefore, since the corners of Bianchi IX asymptotically correspond to the equality of two scale factors, {e.g.,} one is fixed by the condition $\beta_-=0$ and the other two are obtained for the rotational invariance of $2\pi/3$ in the plane $(\beta_+, \beta_-)$, this leads to the Taub solution. The line element of the Taub space-time corresponds to \eqref{linelementbianchi} but the traceless symmetric matrix which determines the anisotropy via $\beta_+$ only is
\begin{linenomath}
\begin{equation}
    \beta_{ab}=diag(\beta_+,\beta_+,-2\beta_+)\,.
\end{equation} 
\end{linenomath}
Within this study, we consider again a cosmological constant $\Lambda$ and a free minimally coupled scalar field $\phi$ to mimic the inflationary scenario. {For further studies about inflation in quantum cosmology, see} \cite{bib:vilenkin-2002, bib:montani-primordialcosmology, bib:weinberg-2008}.
The dynamics is summarized by the scalar constraint
\begin{linenomath}
\begin{equation}
    H_T=K e^{-3\alpha}(-p_{\alpha}^2+p_+^2+p_{\phi}^2+\mathcal{V}+\Lambda e^{6\alpha})=0,
\end{equation}
\end{linenomath}
(we remind that $K=\kappa /3(8\pi)^2$) in which the potential term takes the form
\begin{linenomath}
\begin{equation}
    \mathcal{V}\equiv \frac{3(4\pi)^4}{\kappa^2}\,e^{4\alpha}\,V_T(\beta_+),
\end{equation}
\end{linenomath}
where $V_T(\beta_+)=e^{-8\beta_+}-4e^{-2\beta_+}$. The phase space of the system is six-dimensional with coordinates ($\alpha, p_{\alpha},\beta_+,p_+, \phi, p_{\phi}$) having $p_{\phi}$ as a constant of motion because of the absence of a potential term $U(\phi)$. The dynamical picture is completed by taking into account the choice of $N=e^{3\alpha}/K$ which fixes the temporal gauge.

Now, following the same steps of the Vilenkin approach, we can construct the classical and the quantum dynamics. Equations \eqref{eq:Vilenkin-HJ} and \eqref{eq:VilenkinAmplitude} become
\begin{linenomath}
\begin{equation}
    -(\partial_{\alpha}S)^2+(\partial_{\phi}S)^2+\Lambda e^{6\alpha}=0,
    \label{actionTaub}
\end{equation}
\end{linenomath}
\begin{linenomath}
\begin{equation}
    \partial_{\alpha}(A^2\partial_{\alpha}S)+\partial_{\phi}(A^2\partial_{\phi}S)=0.
    \label{amplitudeTaub}
\end{equation}
\end{linenomath}
Equation \eqref{schro} is instead responsible for the evolution of the quantum subspace, here represented by $\beta_+$: introducing the change of variable $e^{\alpha}=a$, the equation takes the form 
\begin{linenomath}
\begin{equation}
    i\hbar \frac{\partial \chi}{\partial\tau} =\left(-\partial_+^2+\frac{a^4}{4\kappa^2}V_T(\beta_+)\right)\chi,
    \label{schrotaub}
\end{equation}
\end{linenomath}
where $d\tau=K e^{-3\alpha} dt$ and the variable $\alpha$ increases with the synchronous time while $\tau$ decreases. In this respect we have
\begin{linenomath}
\begin{equation}
    \frac{d\alpha}{d\tau}=-2K p_{\alpha}\; < \;0,
    \label{alfatau}
\end{equation}
\end{linenomath}
with $p_{\alpha}\sim \sqrt{\Lambda e^{6\alpha}}$ since in \eqref{actionTaub} $p_{\phi}^2$ can be neglected for large values of $\alpha$. We are also taking the positive square root since we consider an expanding universe.
The behavior of $\tau$ compared to $a$ is then
\begin{linenomath}
\begin{equation}
    \frac{d\tau}{dt}=-\frac{1}{2K \sqrt{\Lambda}a^4} \frac{da}{dt}.
\end{equation}
\end{linenomath}
According to Vilenkin's idea of a small quantum subsystem, the quasi isotropic regime is considered, in which $|\beta_+|\ll 1$, and as a consequence the potential term gets a quadratic form
\begin{linenomath}
\begin{equation}
    V_T(\beta_+)=-3+24\beta_+^2.
\end{equation}
\end{linenomath}
It is worth noticing that the zero order of the approximate potential would provide a contribution to the HJ equation \eqref{actionTaub} and becomes negligible when the cosmological constant dominates, once substituted into the WDW, {i.e.,} $-3e^{4\alpha}\equiv -3a^4$. Hence, the frequency of the harmonic oscillator reads as $\omega^2(\tau)=6\tau^{-4/3}/\tilde{k}^2$ where $\tilde{k}^2=\kappa^2(6\kappa\sqrt{\Lambda})^{4/3}$. Now, with the method used above \cite{bib:lewis-1967}, we can construct an expression for $\rho$ namely
\begin{adjustwidth}{-\extralength}{0cm}
\begin{myequation}
\begin{split}
\rho(\tau)= & \frac{\tilde{k}^3}{324\sqrt{3}} \left\{ \frac{1}{\tilde{k}^2}\left[ (9A^2+64B^2)(\tilde{k}^2+54\ \tau^{2/3}) +\left( (-9A^2+64B^2)(\tilde{k}^2-54\ \tau^{2/3})  -144\sqrt{24A^2B^2-\frac{59049}{k^6}}\tilde{k}\tau^{1/3}\right) \right.\right.\\
&\times \cos\left( \frac{6\sqrt{6}\tau^{1/3}}{\tilde{k}} \right) +6\sqrt{2} \left(2\sqrt{8A^2B^2-\frac{19683}{\tilde{k}^6}}\tilde{k}^2+\sqrt{3}(-9A^2+64B^2)\tilde{k}\tau^{1/3}-108\sqrt{8A^2B^2-\frac{19683}{\tilde{k}^6}} \tau^{2/3} \right) \\
&\left. \left. \hspace{10em} \times \sin\left(\frac{6\sqrt{6} \tau^{1/3}}{\tilde{k}}\right) \right] \right\}^{1/2},
\end{split}
\end{myequation}
\end{adjustwidth}
and the probability density for a generic expansion, {i.e.,} $|\chi(\beta_+,\tau)|^2$, is then calculated.

\begin{figure}[H]
  \includegraphics[width=10.5cm]{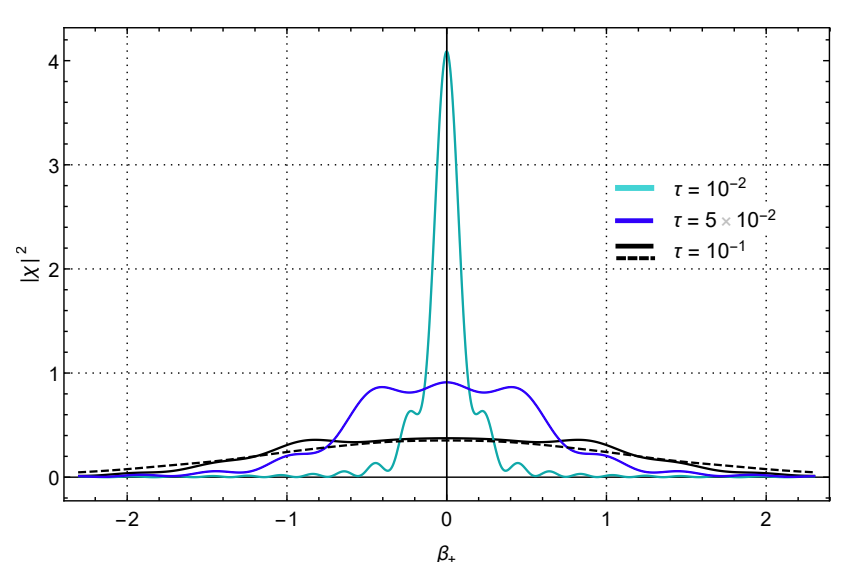}
  \caption{Time evolution of the probability density is highlighted with different colours. The dashed black line represents the initial time $\tau_i$  while the continuous line is the solution with Hermite polynomials. The wavy trend is given due to the truncation of the Hermite polynomials. In this plot we used $A=\frac{81\sqrt{3/2}}{2B}$ and $B=1$. Figure {from} \cite{bib:deangelis-2020}.
}
  \label{taub}
\end{figure}

Figure~\ref{taub} shows that, as the volume of the universe expands, {i.e.,} $\tau \rightarrow 0$, the profile of the Gaussian shape becomes more and more peaked. In this case, it cannot reach a real $\delta$-function as stated in \cite{bib:battisti-belvedere-montani-2009} but a steady small finite value emerges, namely
\begin{linenomath}
\begin{equation}
    \rho(\tau \rightarrow 0)=\frac{2\sqrt{\frac{2}{3}}}{81}\tilde{k}^3+\frac{2}{3}\tilde{k}\tau^{2/3}-\frac{3\sqrt{6}\tau^{4/3}}{\tilde{k}}+\mathcal{O}(\tau)^{5/3}.
    \label{rhotaub}
\end{equation}
\end{linenomath}
A confirmation of this behaviour is present also considering an asymptotic study of an exact Gaussian solution of the time-dependent Schr\"odinger equation, which will be discussed below. We can state that the de Sitter exponential expansion of the universe strongly suppresses the quantum anisotropy leaving a small relic at the end of inflation. This surprising result suggests that, although the anisotropy cannot have the same non-suppressed behaviour of a scalar field, a small tensor degree of freedom can be present on a quantum level. In this sense, in the full inhomogeneous scenario, it could originate a smaller tensorial component of the primordial spectrum. 

\begin{figure}[H]
  \includegraphics[width=10.5cm]{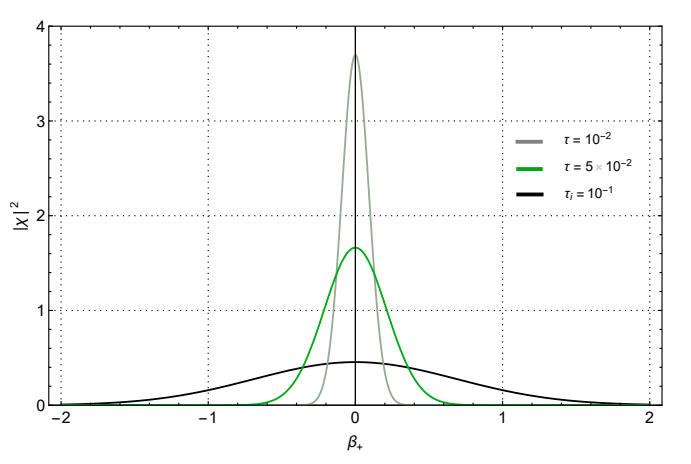}
  \caption{Time evolution of the probability density is highlighted by different colours. The initial time is $\tau_i$. We considered $\frac{1}{f_0}=\frac{2\sqrt{2/3}}{81}$. Figure {from} \cite{bib:deangelis-2020}.}
  \label{taubexact}
\end{figure}

To clarify what we anticipated above, we now search for an exact Gaussian solution~\cite{bib:brizuela-kiefer-2016-desitter,bib:brizuela-kiefer-2016-slow-roll} of the time-dependent Schr\"odinger equation as
\begin{linenomath}
\begin{equation}
    \chi(\beta_+,\tau)=N(\tau)e^{-\frac{1}{2}\Omega(\tau)\beta_+^2},
    \label{chitaub}
\end{equation}
\end{linenomath}
since it is evident from the harmonic oscillator eigenfunction that the simplest way to locate the universe is a Gaussian shape. Substituting \eqref{chitaub} in \eqref{schrotaub} and separating all terms of zero and quadratic order in $\beta_+$, we get
\begin{linenomath}
\begin{equation}
    iN'(\tau)=\frac{1}{2}N(\tau)\Omega(\tau),
\end{equation}
\begin{equation}
    i\Omega'(\tau)=\Omega^2(\tau)-\omega^2(\tau).
    \label{omega}
\end{equation}
\end{linenomath}
To achieve the modulus of the normalization factor we also request a normalized wave function for any value of time.
To obtain the physical information on the anisotropy behaviour we solve \eqref{omega} since the quantity we need is the inverse Gaussian width. Now, separating $\Omega$ into its real and imaginary parts, {i.e.,} $\Omega=f(\tau)+i g(\tau)$, we obtain the following non-linear~system
\begin{linenomath}
\begin{align}
    2g&=\frac{f'}{f}\,, \label{g}\\
    g'&=g^2+\omega^2-f^2.\label{f}
\end{align}
\end{linenomath}
It is worth noting that \eqref{g} and \eqref{f} do not admit an analytical solution, but we can easily construct an asymptotic behaviour for which $\tau \rightarrow 0$. We achieve
\begin{linenomath}
\begin{align}
    g(\tau \rightarrow 0) &\simeq -\frac{3C^2}{\tau^{1/3}},\\
    f(\tau \rightarrow 0) &\simeq f_0\, e^{-6C^2\tau^{2/3}},
\end{align}
\end{linenomath}
where $C^2=6/\tilde{k}^2$ and $f_0$ is an integration constant. The standard deviation of the Gaussian probability distribution is
\begin{linenomath}
\begin{equation}
    \sigma(\tau \rightarrow 0)=\frac{1}{\sqrt{\Re(\Omega)}}\simeq \frac{1}{\sqrt{f_0}}e^{\frac{6}{2}C^2\tau^{2/3}}.
\end{equation}
\end{linenomath}
Hence, the standard deviation exponentially decays (Figure~\ref{taubexact}) when the universe expands, {i.e.,} $\tau \sim 1/a^3$ decreases. However, also here, it approaches a non-zero value. In fact, this feature corresponds to the constant value assumed by $\rho$ in \eqref{rhotaub}. We can state that, if the universe anisotropy is small enough to be in a quantum regime when inflation starts, it is still present at late times.

\subsection{Inhomogeneous Extension}\label{ssec:bianchi-IX-ihnom}

We now briefly review the analysis developed in \cite{bib:montani-chiovoloni-2021}, where the ideas presented above have been extended to the generic inhomogeneous cosmological solution, also clarifying the physical conditions under which the WKB scheme becomes applicable. 

The analysis of a generic inhomogeneous Universe has been first developed in \cite{bib:belinski-k-l-1982}, see also \cite{bib:kirillov-1993,bib:montani-1995,bib:benini-montani-2004,bib:montani-benini-2006} and it corresponds to the situation in which the functions $\alpha$, $\beta_+$ and $\beta_-$ acquire a dependence on the spatial coordinates and the 1-forms, describing the geometry of the 3-hypersurfaces, are associated to a generic vector field, {whose} time dependence is neglected at the higher order.

This scheme allows to implement the so-called ``Belinski–Khalatnikov–Lifshitz (BKL) conjecture'' (for its validation on a classical level see \cite{bib:kirillov-1993,bib:montani-primordialcosmology}), according to which each region of the order of the averaged cosmological horizon behaves like the homogeneous Bianchi IX and Bianchi VIII models, sufficiently close to the initial singularity. In this picture, the chaotic feature of these two Bianchi 
models is extended to the dynamics of a generic inhomogeneous Universe as a local concept: each causal region is characterized by the same oscillatory regime and chaotically evolves independently from any other one. 
Actually, this picture is the result of a more rapid decreasing of the average horizon with respect to the typical inhomogeneous scale, as the initial singularity is approached.
Thus, in the limit of the BKL conjecture validity (for the question concerning possible spikes in the spatial gradients see \cite{bib:uggla-2012-spike}), the Mixmaster scenario described by the triangular potential in Figure~\ref{triangularfig} can be applied as a point-like model, including the corner dynamics addressed above~\cite{bib:montani-primordialcosmology}.

In \cite{bib:montani-chiovoloni-2021} it has been argued that, inside the corner, the variables $\alpha$ and $\beta_+$ are, near the singularity, very large and therefore remain classical degrees of freedom, while the small variable $\beta_-$ can become a real quantum variable, according to the proposal in \cite{bib:vilenkin-1989}. The idea is that, in a long sequence of iterations of the piecewise representation of the evolution in terms of the Kasner-like solution, a deep penetration of the point-universe inside the corner must, soon or later, take place in each spatial point \cite{bib:lifshitz-al-1985}.

By other words, it is argued that the uncertainty in the value of the variable $\beta_-$ in the corner is of the order $\Delta \beta_- \sim 2\beta_+\sqrt{\hbar}$. 
Hence the uncertainty principle implies that the indetermination on the corresponding momentum is $\Delta p_- \sim 2\sqrt{\hbar}/\beta_+$. 
Recalling that deeply in the corner $\beta_+$ is very large, we deal with a small quantum subsystem associated to the phase space $\{ \beta_-,\, p_-\}$ and the ratio between the quantum Hamiltonian and the classical one is of order $\hbar$. That is, all the assumptions at the ground of the decomposition into two parts of the global system, one classical (here the quantum corrections on $\alpha$ and $\beta_+$ are not present at all) and a small quasi-classical subset, considered in \cite{bib:vilenkin-1989} are fully satisfied.
Hence, the same analysis performed above follows directly in each space point, since the variable $\beta_-$ dynamics is descried by a time-dependent quantum harmonic oscillator in each locally homogeneous region. However, in the inhomogeneous case, it has to be taken into account the so-called fragmentation of the space \cite{bib:belinski-1992,bib:montani-1995,bib:barrow-2020,bib:montani-primordialcosmology}. In fact, the chaotic time evolution of the locally homogeneous regions induces a corresponding oscillation of the spatial dependence of the metric functions. As a result, the comoving inhomogeneity scale is not the same during all the evolution toward the singularity, but it also decreases~\cite{bib:kirillov-1993,bib:montani-primordialcosmology}, although the Mixmaster scenario is preserved. 

The important point here is that the corner configuration is then reached in each (even arbitrarily small) space region which contains a rational value of the parameter $u$, by which the BKL map is described \cite{bib:belinski-k-l-1982,bib:montani-1995}.
This feature ensures that, as the initial singularity is approached, essentially all the space is (homogeneous patch by homogeneous patch) in the corner configuration (this takes place in different instants of time) and the WKB scenario inferred in Section~\ref{sec:vilenkin} can be applied. 
Thus, in the end, since the variable $\beta_-$ is frozen out to a negligible value (described by a constant standard deviation around the zero mean value), then we deal with a non-singular generic inhomogeneous universe. 
In this respect, the implementation of the ideas developed in \cite{bib:vilenkin-1989} to the inhomogeneous Mixmaster leads to a possible picture to solve the problem of the initial singularity on a very general footing.

\section{WKB Expansion for Quantum Gravity Contributions}\label{sec:WKB}

As seen in Section~\ref{sec:vilenkin}, the work \cite{bib:vilenkin-1989} implemented a perturbative expansion in the Planck constant in order to obtain a functional probabilistic interpretation for the wave function of the universe. This procedure can be enclosed as a special case of the WKB method \cite{bib:dunham-1932}, which allows to compute an approximate solution to a differential equation of the WDW type by going to increasing orders of accuracy in a desired parameter, as seen in the study presented in Section~\ref{sec:appl-vilenkin-bianchiI}.
To illustrate this method, let us start from the WDW equation \eqref{eq:WDW}; we assume the solution wave function $\Psi$ to be of the form
\begin{linenomath}
\begin{equation}\label{eq:psiWKB}
	\Psi = e^{i S/ \hbar} \,,
\end{equation}
\end{linenomath}
with $S$ a complex function, that we expand in some parameter $P$
\begin{linenomath}
\begin{equation}\label{eq:espansioneSWKB}
	S = \sum_{n = 0}^{\infty} P^n S_n .
\end{equation}
\end{linenomath}
The application of the superHamiltonian operator to \eqref{eq:psiWKB} using \eqref{eq:espansioneSWKB} brings a series of equations, each one at a different order in $P$ and acting as a small perturbation to the previous ones (having chosen $P$ appropriately).
Using the Planck constant $\hbar$ as expansion parameter \cite{bib:landau-quantumMechanics} this corresponds to the so-called \emph{semiclassical approximation} in quantum theory; nonetheless, one could implement a different perturbation parameter according to the physical properties of the considered theory.
Substituting \eqref{eq:psiWKB} into \eqref{eq:WDW}, one can solve each order in $P$ and, supposing that the universe can be separated as in Section \ref{sec:vilenkin}, obtain a dynamical description of the quantum subsystem at some level of accuracy, containing corrections from the \lq\lq semiclassical'' sector.

We emphasize the difference between the direct application of this method and the work \cite{bib:vilenkin-1989} presented in Section \ref{sec:vilenkin}, {i.e.,} the ansatz \eqref{eq:defVilenkinSemiclassical} was composed of a semiclassical amplitude $A(h)$, multiplied by an exponential term expanded in $\hbar$. This hypothesis is based on the assumption that the universe can, at some level, be separated between a purely semiclassical sector and the remaining quantum one, as already discussed, but it is not a general feature of the WKB method. However, Vilenkin's work can be recast as a WKB expansion with the ansatz \eqref{eq:psiWKB} by considering a complex function $S$ and expanding~\eqref{eq:espansioneSWKB} in the parameter $\hbar$, as shown in \cite{bib:montani-digioia-maniccia-2021}.

The WKB expansion for quantum gravity has been implemented in many works in the literature after \cite{bib:vilenkin-1989}, mainly focusing on the canonical quantization prescription \cite{bib:lifschytz-1996,bib:castagnino-1993,bib:barbour-1994,bib:okhuwa-1995,bib:damour-vilenkin-2019} sometimes in different expansion parameters \cite{bib:castagnino-1990,bib:kiefer-1991,bib:moffat-1993,bib:barvinsky-kiefer-1998,bib:bolotin-2015,bib:brizuela-kiefer-2016-desitter,bib:brizuela-kiefer-2016-slow-roll,bib:kiefer-2018,bib:giulini-2012,bib:kiefer-2019,bib:rotondo-2020,bib:rotondo-2022}, or taking different paths considering some sort of WKB ansatz \cite{bib:halliwell-1989,bib:barvinsky-1990,bib:barbour-1993,bib:robles-perez-2021}. In several works, the WKB method has been implemented in the context of a BO approximation \cite{bib:born-1927,bib:bransden} for gravity and matter \cite{bib:bertoni-venturi-1996,bib:massar-1998,bib:venturi-2017,bib:venturi-2020,bib:chataignier-2020,bib:venturi-2021,bib:chataignier-2021,bib:montani-digioia-maniccia-2021,bib:maniccia-montani-2022,bib:chataignier-2022}. Reviews regarding the use of the WKB procedure for constructing time in quantum cosmology can be found in \cite{bib:unruh-wald-1989,bib:kiefer-1994-review,bib:kuchar2011-review,bib:kiefer-2013-review,bib:peter-kiefer-2022}. In the following section we will explore in more detail some of these works and discuss the emerging problem of non-unitarity for the matter dynamics.

\subsection{Time from Gravitational Variables and the Question of Non-Unitarity}\label{ssec:non-unitarity}

The expansion parameter in \eqref{eq:espansioneSWKB} can also be taken of Planckian size. That is the case of Kiefer and Singh's work \cite{bib:kiefer-1991}, who first considered a regime in which the \lq\lq classical limit'' is the absence of matter, {i.e.,} vacuum solutions.

Let us briefly recall this approach. We start by identifying in the system the \lq\lq subsets'' of quantum gravity and quantum matter, such that the WDW equation can be rewritten as
\begin{linenomath}
\begin{equation}\label{eq:WDWkiefer}
    \left( - \frac{\hbar^2}{2 M}\left( \nabla_g^2 + f\cdot \nabla_g \right)  + M V (g) + \hat{H}_m \right) \Psi (g, m) = 0 \,,
\end{equation}
\end{linenomath}
where $M$ is the Planckian parameter
\begin{linenomath}
\begin{equation} \label{eq:defM}
    M \equiv \frac{1}{4 c^2 \kappa} = \frac{c m_\mathrm{P}^2}{4 \hbar} ,
\end{equation}
\end{linenomath}
being $m_P = \sqrt{\hbar c / 8\pi G}$ the reduced Planck mass, the term $f\cdot \nabla_g$ is inserted for generalization to other operator orderings, and $H_m$ is the (scalar) matter superHamiltonian as in \eqref{eq:hamiltonian-inflaton}. An important aspect deriving from the choice of the expansion parameter \eqref{eq:defM} is that it allows a clear separation between the gravitational and matter subsets, since in the limit $M \rightarrow \infty$ ($ G \rightarrow 0 $ as can be seen from \eqref{eq:defM}) the latter will disappear, leaving only the Einstein's equations in vacuum. Such a choice implies that the WKB expansion will hold for particles with small mass over Compton length ratio, {i.e.,} whose mass is $m \ll m_P$.

Similarly to Section~\ref{sec:vilenkin}, the wave function is taken to be of the WKB form
\begin{linenomath}
\begin{equation}
    	\Psi (g,m) = e^{\frac{i}{\hbar} S(g,m)} ,
\end{equation}
\end{linenomath}
and $S$ is then expanded in powers of $M$. However, in Vilenkin's work \cite{bib:vilenkin-1989}, the study was carried out to recover a Schr\"{o}dinger dynamics for the quantum (here matter) variables, and to formulate a probabilistic interpretation for the complete $\Psi$, for which the order $\hbar^1$ was enough. In \cite{bib:kiefer-1991} instead, the aim is not only to recover such a dynamics for the matter sector (which will emerge at $\ord{M^0}$), but also to investigate its modifications induced by the quantum nature of gravity, {i.e.,} going up to the next order $\ord{M^{-1}}$. To obtain this, the total function $S$ is first expanded in powers of $M$ and then at each order separated in $a(g) + b(m,g)$, {i.e.,} isolating a purely gravitational function. For the sake of clarity, we here reformulate the approach with that separation from the beginning, writing
\begin{linenomath}
\begin{equation}\label{eq:ansatz-S-kiefer}
    S (g,m) = M S_0(g) + S_1(g) + \frac{1}{M} S_2(g) + Q_1(m,g) +\frac{1}{M} Q_2(m,g) + \ord{M^{-2}},
\end{equation}
\end{linenomath}
where for consistency the highest function $S_0$ at $\ord{M}$ (Planck scale) depends on gravitational variables only, as can be checked from the perturbative expansion. The matter enters at the next order, such that the gravitational background is naturally recovered without further assumptions. This feature represents a striking difference from the work in Section~\ref{sec:vilenkin}, where the WDW gravitational equation was also imposed. We stress that, in this implementation, the presence of classical matter can only be recovered with some suitable redefinition, for example with a rescaling of the matter fields themselves (see \cite{bib:brizuela-kiefer-2016-slow-roll,bib:brizuela-kiefer-2016-desitter}).

Expanding in $M$, the first order $M^1$ gives
\begin{linenomath}
\begin{equation}
    \label{eq:HJ_M}
	\frac{1}{2} (\nabla_g S_0)^2 + V = 0,
\end{equation}
\end{linenomath}
corresponding to the HJ equation for gravity which provides the classical limit, namely Einstein's equations in vacuum. We note that the coefficient $1/2$ in front of $(\nabla_g S_0)^2$ with respect to Vilenkin's proposal \eqref{eq:Vilenkin-HJ} is due to the definition of the expansion parameter $M$ which makes it appear in the starting WDW Equation \eqref{eq:WDWkiefer}. In this sense, it is not related to any physical properties.
The next order $M^0$ brings
\begin{adjustwidth}{-\extralength}{0cm}
\begin{equation}
    	\nabla_g S_0 \cdot \nabla_g S_1 + \nabla_g S_0 \cdot \nabla_g Q_1 - \frac{i\hbar}{2} \left( \nabla_g^2 S_0 + f\cdot \nabla_g S_0 \right) +\frac{1}{2\sqrt{h}} (\nabla_m Q_1)^2 -\frac{i\hbar}{2\sqrt{h}} \nabla_m^2 Q_1 + U= 0\,,
\end{equation}
\end{adjustwidth}
where we indicate the derivatives with respect to $\phi$ as $\nabla_m$. Requiring that $S_1(g)$ satisfies
\begin{linenomath}
\begin{equation}\label{eq:kiefer-continuity-S1}
     \nabla_g S_0 \cdot \nabla_g S_1 -\frac{i\hbar}{2}\left( \nabla_g^2 S_0 + f\cdot \nabla_g S_0 \right) =0 \,,
\end{equation}
\end{linenomath}
namely a continuity equation for $S_1$ (being $S_0$ known from the previous order), the matter wave function $\chi_0 = e^{\frac{i}{\hbar} Q_1}$ satisfies\vspace{6pt}
\begin{linenomath}
\begin{equation}\label{eq:schrod-kiefer}
    i\hbar \dtau{} \chi_0 = N \hat{H}_m \,\chi_0\,.
\end{equation}
\end{linenomath}
Equation~\eqref{eq:schrod-kiefer} is a functional Schr\"{o}dinger equation where the WKB time is defined by
\begin{linenomath}
\begin{equation}\label{eq:deftimeKiefer}
    \dtau{} = N \,\nabla_g S_0 \cdot \nabla_g,
\end{equation}
\end{linenomath}
similar to \eqref{eq:deftimeVilenkin}, in which the lapse function (that was removed in the original work via a gauge choice) has been reinserted for the general case in order to maintain a parallelism with Vilenkin's definition \eqref{eq:deftimeVilenkin}. We emphasize that, in Section~\ref{sec:vilenkin}, the continuity equation was not imposed but obtained from the perturbative procedure since we required the WDW gravitational constraint from the beginning; here instead, there is no such initial assumption. To recover the functional quantum field dynamics we have to impose another condition on $S_1$, {i.e.,} \eqref{eq:kiefer-continuity-S1}.

Developing the analysis to the next order $M^{-1}$, one finds
\begin{adjustwidth}{-\extralength}{0cm}
\begin{equation}
\begin{split}
    \nabla_g S_0 \cdot \nabla_g S_2 + \nabla_g S_0 \cdot \nabla_g Q_2 +& \frac{1}{2} \left( (\nabla_g S_1)^2  +(\nabla_g Q_1)^2 \right)+ \nabla_g S_1 \cdot \nabla_g Q_1 -\frac{i\hbar}{2} \left( \nabla_g^2 S_1 + \nabla_g^2 Q_1 + f\cdot \nabla_g S_1 \right.\\
    &\left.\vphantom{\nabla_g^2}\quad+ f\cdot \nabla_g Q_1 \right) + \frac{1}{\sqrt{h}} \nabla_m Q_1 \nabla_m Q_2 -\frac{i\hbar}{2\sqrt{h}} \nabla_m^2 Q_2 =0 \,,
    \end{split}
\end{equation}
\end{adjustwidth}
which again can be cast in a clearer form once the function $S_2$ satisfies an analogous continuity equation
\begin{linenomath}
\begin{equation}
    \nabla_g S_0 \cdot \nabla_g S_2 + \frac{1}{2}(\nabla_g S_1)^2 -\frac{i\hbar}{2} \left( \nabla_g^2 S_1 +f\cdot \nabla_g S_1 \right) =0,
\end{equation}
\end{linenomath}
thus leaving only
\begin{adjustwidth}{-\extralength}{0cm}
\begin{equation}\label{eq:kieferM-1}
    \nabla_g S_0 \cdot \nabla_g Q_2 +\frac{1}{2} (\nabla_g Q_1)^2 + \nabla_g S_1 \cdot \nabla_g Q_1 -\frac{i\hbar}{2} (\nabla_g^2 Q_1 +f\cdot \nabla_g Q_1) + \frac{1}{\sqrt{h}} \nabla_m Q_1 \nabla_m Q_2 -\frac{i\hbar}{2\sqrt{h}} \nabla_m^2 Q_2 =0.
\end{equation}
\end{adjustwidth}
We can now decompose the derivatives $\nabla_g$ in tangent and normal components to the hypersurfaces $S_0 = const$ and neglect the former by assuming the adiabatic dependence of $H_m$ on the induced metric. Summing \eqref{eq:kieferM-1} with the previous order, the resulting equation for the matter wavefunction $\chi = e^{\frac{i}{\hbar} \left(Q_1 + \frac{1}{M} Q_2 \right)}$ for $N=1$ is
\begin{linenomath}
\begin{equation}\label{eq:Kieferfinal}
    i\hbar \dtau{\chi} = \hat{H}_m \chi +\frac{1}{8M\, \sqrt{h} \bar{R}} \left[ \hat{H}_m^2 +i\hbar \left( \dtau{H_m} -\frac{1}{\sqrt{h}\bar{R}} \dtau{(\sqrt{h}\bar{R})} \hat{H}_m \right) \right] \chi \,.
\end{equation}
\end{linenomath}
Here, the terms after $H_m$ are a modification to the standard quantum matter dynamics and thus they represent quantum gravity corrections. An inspection of these terms reveals that they violate unitarity in the evolution.

It can be noted that, up to the order $M^0$, the work \cite{bib:kiefer-1991} seems to portray a functional description of the system analogous to the one obtained by Vilenkin (Section~\ref{sec:vilenkin}). Actually, it can be shown that the approaches \cite{bib:vilenkin-1989,bib:kiefer-1991} are equivalent to a unique WKB expansion of the WDW equation just by changing the expansion parameter (see reformulation in \cite{bib:montani-digioia-maniccia-2021}). As a consequence, Vilenkin's work can also be expanded to the next order in $\hbar$ finding quantum gravity corrections in the functional Schr\"{o}dinger formalism. However, also in that case, they manifest a non-unitary morphology.

The question of non-unitarity in this kind of approaches has been long discussed in the literature \cite{bib:barvinsky-1990,bib:barvinsky-1993,bib:bertoni-venturi-1996,bib:mostafa-2004,bib:kiefer-2018,bib:chataignier-2021,bib:montani-digioia-maniccia-2021,bib:gielen-2022}, {with many significant outcomes. As presented in}~\cite{bib:gielen-2022}, {implementing a scalar field clock, the request of unitarity can lead to a quantum recollapse of the model; in}  \cite{bib:chataignier-2021} {an inner product is proposed in relation to the Faddeev--Popov gauge-fixing procedure.} {We here briefly discuss the proposal} \cite{bib:kiefer-2018} to overcome the non-unitarity emerging in Equation~\eqref{eq:Kieferfinal}: the authors construct the set of complex eigenvalues $E(\tau)$ associated to the total non-Hermitian Hamiltonian operator in \eqref{eq:Kieferfinal}, together with the set of real eigenvalues $\epsilon(\tau)$ of $\hat{H}_m$. In this notation, $\tau$ is the only geometrical variable present that is identified as time from the beginning. The functions $E(\tau)$ and $\epsilon(\tau)$ are then expanded in powers of $1/M$. By redefining the quantum wave function with a phase transformation involving the imaginary part of $E(\tau)$, and rescaling the background with the opposite phase, the redefined quantum state gives a contribution in the equation that exactly cancels the non-unitary terms in \eqref{eq:Kieferfinal}. Thus, the dynamics for the redefined $\chi$ at $\ord{M^{-1}}$ presents only the Hermitian part of the quantum gravity corrections, restoring unitarity; also, a quantum backreaction emerges in the HJ equation due to the rescaling. However, the procedure is built on the assumption that the operators $H_{tot}$ and $H_m$ commute, and thus can be diagonalized simultaneously. This property does not hold in some cases, for instance considering a FLRW model with a cosmological constant and a scalar field. In that setting, $H_{tot}$ at the order $1/M$ contains both $H_m$ and its time derivative $\dot{H}_m$, with $H_m$ including the scale factor $a$ and $\dot{H}_m$ its conjugate momentum, so the two operators cannot commute (for a critical analysis of this restatement, see \cite{bib:montani-digioia-maniccia-2021}).

{Moreover, the question of non-unitarity has been addressed also in the context of modified theories of gravity, where it can emerge due to renormalizability requirements of the corresponding quantum theory (e.g.,} \cite{bib:fradkin-1981}). {Recent interest has been devoted to the case of massive gravity, where the graviton particle acquires a nonzero mass. Massive gravity was first introduced by the work of Pauli and Fierz} \cite{bib:fierz-pauli-1939} {and later reformulated with the \lq\lq gravitational Higgs mechanism'' (in which the spontaneously broken symmetry is the one associated to coordinate reparametrization invariance) or via higher-derivative curvature terms}  \cite{bib:hinterbichler-2012}. {Such theory is however plagued by the emergence of ghost fields, i.e., non-physical states associated to non-dynamical variables, that induce negative probabilities in the theory and so violate unitarity} \cite{bib:boulware-deser-1972,bib:creminelli-2005}. {Solutions to this issue have been proposed both in three dimensions, see} \cite{bib:bergshoeff-2009,bib:masashi-oda-2009,bib:arvanitakis-2015,bib:setare-2015} {and in four dimensions with the so-called dRGT model}~\cite{bib:de-rham-2011} {(see} \cite{bib:arraut-2015} {for some deviations from GR predicted by the model), and \mbox{also~\cite{bib:park-2011,bib:paulos-2012,bib:einhorn-2017}}}. 

{For what concerns the non-unitarity problem in the present General Relativity analysis,} the description of quantum gravity corrections to the matter sector dynamics with the WKB procedure leaves some unanswered questions. Another relevant implementation is to regard the gravity and matter system in a Born--Oppenheimer approximation, as mentioned in Section~\ref{sec:appl-vilenkin-bianchiI}, in order to tackle this issue in the canonical quantization framework.

\subsection{The Born--Oppenheimer-like Approximation}\label{ssec:BO}

A further implementation of the DeWitt theory for gravity and matter is the Born--Oppenheimer (BO) extended approach presented in \cite{bib:bertoni-venturi-1996}, later applied in the context of quantum cosmology in \cite{bib:venturi-2017,bib:venturi-2020,bib:venturi-2021}. In analogy with the BO approximation for molecules, the wave function is separated as $\Psi (g,m) = \psi (g) \chi(m,g)$ since the matter sector is characterized by a lower mass scale with respect to the Planckian one. Hence, the matter can be regarded as the \lq\lq fast'' quantum sector while gravity is the \lq\lq slow'' quantum component. Working in the minisuperspace, the total WDW equation \eqref{eq:WDW} is averaged over $\chi(m,g)$ and subtracted to the initial equation thus obtaining an equation for the gravitational background $\psi$ and one for the matter sector $\chi$. Both functionals are rescaled making use of the gauge invariance of the system through a phase depending only on the gravitational~variables
\begin{linenomath}
\begin{equation}
    \psi = e^{- \frac{i}{\hbar} \int A \,dg} \widetilde{\psi}, \quad \chi = e^{\frac{i}{\hbar} \int A\,dg} \widetilde{\chi},
\end{equation}
\end{linenomath}
where $	A = - i\hbar \langle \nabla_g \rangle$. Then, rescaling again $\chi$ via $\langle H_m \rangle$ and taking $\psi$ in the WKB form, the HJ Equation \eqref{eq:HJ_M} is modified by the presence of the matter backreaction $\langle H_m \rangle$.  Implementing the time definition \eqref{eq:deftimeKiefer}, the dynamics of the matter sector is given by
\begin{linenomath}
\begin{equation}\label{eq:schrodmodifiedVenturi}
    \left( \hat{H}_m - i\hbar \dtau{}\right) \chi_s = e^{- \frac{i}{\hbar} \int \langle H_m \rangle \,d\tau - \frac{i}{\hbar} \int A \,dg} \,\frac{\hbar^2}{2 M} \left[ \bar{D}^2 -\langle \bar{D}^2 \rangle + 2 ( D \ln \mathcal{N} ) \bar{D} \right] \chi \,,
\end{equation}
\end{linenomath}
where $D$, $\bar{D}$ are covariant derivatives constructed with $A$ as Berry connection, $1/\mathcal{N}$ is the amplitude associated to the WKB-expanded $\psi$, and $\chi_s = e^{- \frac{i}{\hbar} \int \langle H_q \rangle \,d\tau - \frac{i}{\hbar} \int A \,dg} \chi $. As in the previous approaches, in the semiclassical limit the right-hand side vanishes due to the adiabatic approximation and Equation~\eqref{eq:schrodmodifiedVenturi} describes the usual Schr\"{o}dinger dynamics. Furthermore, the authors suggest that the obtaining dynamics is unitary due to the vanishing of
\begin{linenomath}
\begin{equation}\label{eq:VenturiUnitarityCondition}
    i \hbar \dtau{}\langle \chi_s | \chi_s \rangle = 0 \,.
\end{equation}
\end{linenomath}

However, this approach does not completely solve the non-unitarity problem. In fact, while the norm of quantum states preserves unitarity signaling a possible construction of the Hilbert space associated to the matter sector, this might not be true when the quantity~\eqref{eq:VenturiUnitarityCondition} is computed between different quantum states. It has also been shown in \cite{bib:montani-digioia-maniccia-2021} that, once the gravitational wavefunction $\psi$ is rescaled with $\langle H_m \rangle$ (which is a requirement of the gauge symmetry of the theory), Equation~\eqref{eq:schrodmodifiedVenturi} takes a different form, again as a modified Sch\"{o}dinger equation that is unitary only if one considers $\langle \chi_s | \chi_s \rangle$. Moreover, as a consequence of the rescaling, the matter backreaction does not appear at the level of the HJ but goes to the next order where it gets canceled by an opposite term, actually vanishing in the proposed approach. 

\textls[-10]{The presence of the quantum backreaction in these models is also worth discussing~\cite{bib:schander-thiemann-2021}}. Considering Vilenkin's work, this contribution is absent from the HJ due to the background assumption \eqref{eq:Vil-WDW-semicl}, while in \cite{bib:kiefer-1991} it is forbidden by the choice of expansion parameter, as mentioned above. However, using the same parameter, a matter backreaction term emerges in both \cite{bib:kiefer-2018,bib:bertoni-venturi-1996} via some rescaling. In the context of quantum cosmology, when perturbations are present, such backreaction would describe how small scale inhomogeneities influence the large-scale structure of the universe. With this aim, many studies have been carried on considering both semiclassical and quantum backreactions, {i.e.,} with a classical or quantized gravitational sector (see \cite{bib:schander-thiemann-2021} and references within for an overview). In relation to the topics here presented, we mention the implementations based on Space-Adiabatic Perturbation Theory (SAPT) \cite{bib:panati-2002-sapt}, which can be formulated as a generalization of the Born--Oppenheimer procedure aimed at solving the coupled dynamics at a perturbative level \cite{bib:thiemann-2016-sapt,bib:thiemann-2019-sapt}.

\section{A Proposal for Unitarity: The Role of the Reference System}\label{sec:kin-fluid-unitarity}

The emergence of non-unitarity in the approaches discussed above may signal that the time definitions in \eqref{eq:deftimeVilenkin}, \eqref{eq:deftimeKiefer} are to be reconsidered. Indeed, they bring in the expansion at $\ord{M^{-1}}$ (or $\ord{\hbar}$ in Vilenkin's approach) a squared time derivative coming from $\nabla_g^2$ which leads to non-unitary terms in the modified dynamics \cite{bib:montani-digioia-maniccia-2021}. 

A different implementation of time can follow from exploiting the role of the reference frame, whose presence in the model can be made explicit by adding a suitable term to the action. In the following we will focus on two different types of this implementation, namely the kinematical action and the Gaussian reference frame fixing, discussing their relation and physical meaning.

\subsection{The Kinematical Action Proposal}\label{ssec:kin-action}
Let us first review Kuchar's discussion presented in \cite{bib:kuchar-1981}. There, the \emph{kinematical action} is defined as the term to be added to the theory, using some Lagrange multipliers, to restore covariance under the ADM foliation and thus under the choice of reference frame. This procedure stems from the observation that, in quantum field theory with an assigned ADM foliation, the relation between points on infinitesimally close hypersurfaces is not evident, {i.e.,} the geometrical meaning of the deformation vector and its components $N$ and $N^i$ is lost, as can be seen in the case of a scalar matter field theory \cite{bib:kuchar-1981,bib:montani-2002,bib:montani-digioia-maniccia-2021}.
In the ADM representation, the kinematical action takes the form\vspace{6pt}
\begin{linenomath}
\begin{equation}\label{eq:defSkin}
    S^{kin} = \int dt\, d^3x (p_{\mu} \partial_t y^{\mu} - N^{\mu} p_{\mu}),
\end{equation}
\end{linenomath}
where $y^{\mu} = y^{\mu} (x^i; x^0)$ define the family of one-parameter hypersurfaces obtained via the foliation, and $p_{\mu}$ are conjugate to $y^{\mu}$. Adding \eqref{eq:defSkin} to the action of the model, further equations of motion (associated to the variations $\delta y^{\mu}$, $\delta p_{\mu}$ and $\delta N^{\mu}$) describe the vanishing of the momenta $p_{\mu}$ and restore the geometrical definition of the deformation vector
\begin{linenomath}
\begin{equation}
	\label{eq:defVector}
	N^{\mu} = \partial_t y^{\mu} = N n^{\mu} + N^i b_i^{\mu} \,,
\end{equation}
\end{linenomath}
being $n^{\mu}$ the timelike direction and $b_i^{\mu}$ the tangent basis to the hypersurfaces identified by the foliation. The superspace constraints are modified by the presence of 
\begin{linenomath}
\begin{gather}
	H^{kin} = n^{\mu} p_{\mu} \label{eq:defHkin}\,,\\
	H_i^{kin} = b_i^{\mu} p_{\mu} \label{eq:defHikin} \,,
\end{gather}
\end{linenomath}
such that the total superHamiltonian and supermomentum functions must now vanish. We notice that these terms represent a good candidate for the definition of time since Equations~\eqref{eq:defHkin} and \eqref{eq:defHikin} are linear in the momenta $p_{\mu}$. 

Let us now analyze the model following from the definition of time through the kinematical action, as implemented in \cite{bib:montani-digioia-maniccia-2021}, instead of background variables. Starting from the action
\begin{adjustwidth}{-\extralength}{0cm}
\begin{equation}
    S^{g} + S^{m} + S^{kin} = \int d x^0 \, d^3 x \left[ \Pi_{a} \dot{h}^{a} + p_{\mu} \dot{y}^{\mu} + \pi \dot{\phi} - N \left(H^{g} + H^{m} + H^{kin}\right) - N^i \left(H^{g}_i + H^{m}_i + H^{kin}_i \right) \right],
\end{equation}
\end{adjustwidth}
and separating the wave function in $\Psi ( h, \phi, y^{\mu} ) = \psi (h) \chi (\phi, y^{\mu} ; h)$ as in Section~\ref{ssec:BO}, the WKB expansion in the Planckian parameter $M$ \eqref{eq:defM} can be performed
\begin{linenomath}
\begin{equation}\label{eq:ansatz-azione-cin}
    \Psi (h, \phi, y^{\mu}) = e^{ \frac{i}{\hbar} \left(M S_0 + S_1 + \frac{1}{M} S_2\right) } \, e^{\frac{i}{\hbar} \left( Q_1 +\frac{1}{M} Q_2\right)}\,,
\end{equation}
\end{linenomath}
being $S_n = S_n(h)$ and $Q_n = Q_n(\phi,y^{\mu};h)$. We stress that, in this separation, the kinematical action (and so the reference frame) is enclosed in the fast quantum sector as are the matter fields, in contrast with the gravitational background; this requirement allows the time parameter to be independent from slow background variables which are related to non-unitarity. In Equation~\eqref{eq:ansatz-azione-cin}, as in \eqref{eq:ansatz-S-kiefer}, the expansion is truncated at order $M^{-1}$ since the aim is to compute quantum gravity corrections to the matter dynamics.
The requirements
\begin{linenomath}
\begin{gather}
    \frac{\langle\hat{H}^m \chi\rangle}{\langle\hat{H}^g \Psi\rangle} = \ord{M^{-1}}\,,\label{eq:Skin-condition-Hsmall}\\ \frac{\delta}{\delta h_{ij}} Q_n(\phi, y^{\mu}; h) =\ord{M^{-1}}, \label{eq:Skin-condition-slow-derivatives}
\end{gather}
\end{linenomath}
are satisfied due to the difference in physical scales and in ``velocities'' of the two sectors typical of the BO approximation, as discussed in Section~\ref{ssec:BO}. Following Vilenkin's reasoning, the total WDW equation is imposed together with the analogous equation for the gravitational background, {i.e.,}
\begin{linenomath}
\begin{gather}
    \left[ -\frac{\hbar^2}{2M} \left( \nabla_g^2 + f \cdot \nabla_g \right) + M V(g) -\hbar^2 \nabla_m^2 + U - i \hbar \, n^{\mu} \frac{\delta}{\delta y^{\mu}}\right] \Psi =  0\,,\label{eq:Skin-superHtot}\\
    \left[ -\frac{\hbar^2}{2M} \left( \nabla_g^2 + f \cdot \nabla_g\right) + MV(g) \right] \psi = 0 \,, \label{eq:Skin-superHgrav}
\end{gather}
\end{linenomath}
where the term $f\cdot \nabla_g$ has been introduced for generic operator orderings, as in Section~\ref{ssec:non-unitarity}, the matter sector is described by a scalar field $\phi$ and the gravitational sector potential $V$ possibly includes a cosmological constant term. In the general case, one cannot implement the minisuperspace reduction, thus the theory must take into account also the supermomentum constraints for the total $\Psi$ and for the background respectively
\begin{linenomath}
\begin{gather}
    \left[ 2h_{i}\,\bar{D} \cdot \nabla_g - \partial_i \phi \cdot \nabla_m  - i\hbar \, b^{\mu}_i \frac{\delta}{\delta y^{\mu}}\right] \Psi = 0 \,,\label{eq:Skin-supermtot}\\
    [2 i\hbar \, h_{i} \bar{D}\cdot \nabla_g ]\psi = 0 \,,\label{eq:Skin-supermgrav}
\end{gather}
\end{linenomath}
being $h_i\, \bar{D} \cdot \nabla_g = h_{ij} \bar{D}_k \frac{\partial}{\partial h_{kj}}$ and $\bar{D}_k$ the (3-dimensional) induced covariant derivative associated to $h_{ij}$. We stress that, since we are here presenting the more general formalism, $i,j,k$ are explicited spatial indices; we will then implement and discuss the minisuperspace reduction of this model.

Substituting \eqref{eq:ansatz-azione-cin}, the expansion of the constraints Equations \eqref{eq:Skin-superHtot}--\eqref{eq:Skin-supermgrav} brings at $\ord{M}$
\begin{linenomath}
\begin{subequations}
\begin{gather}	
	\frac{1}{2} \nabla_g S_0 \cdot \nabla_g S_0 + V = 0\,,\label{eq:Skin-HJ}\\
	-2 h_{k} \bar{D} \cdot \nabla_g S_0 = 0 \,,\label{eq:Skin-diffeom-S0}
\end{gather}
\end{subequations}
\end{linenomath}
corresponding to the HJ and the diffeomorphism invariance of $S_0$. At $\ord{M^0}$, from the gravitational constraint, we obtain a relation between $S_0$ and $S_1$. Using this link and summing Equations~\eqref{eq:Skin-superHtot} and \eqref{eq:Skin-supermtot} with coefficients $N$ and $N^i$ respectively, one obtains
\begin{linenomath}
\begin{equation}\label{eq:Skin-schrod-time}
    i\hbar \dtau{\chi_0} \equiv i\hbar \int d^3 x \left(N n^{\mu} + N^i b^{\mu}_i\right) \frac{\delta }{\delta y^{\mu}} \chi_0 = \hat{\mathcal{H}}^m \chi_0 = \int d^3 x \left(N \hat{H}^m + N^i \hat{H}^m_i\right) \chi_0,
\end{equation}
\end{linenomath}
where $\chi_0 = e^{\frac{i}{\hbar}Q_1}$ is the matter wavefunction at $\ord{M^0}$ and the time derivative, which is defined via the kinematical momenta $p_{\mu}$, includes the definition of the deformation vector $N^{\mu}$. At the next order $M^{-1}$, proceeding in a similar way and making use of the hypothesis~\eqref{eq:Skin-condition-slow-derivatives}, the modified matter dynamics is obtained
\begin{linenomath}
\begin{equation}\label{eq:final-kin-action}
    i\hbar \dtau{\chi} = \hat{\mathcal{H}}^m \chi + \int d^3 x \left[ N \nabla_g S_0 \cdot \left(-i\hbar \nabla_g\right) -2 N^k h_{k} \bar{D} \cdot \left(-i\hbar \nabla_g \right) \right] \chi \,,
\end{equation}
\end{linenomath}
being $\chi = e^{\frac{i}{\hbar}\left(Q_1 +\frac{1}{M} Q_2\right)}$. We can observe that the quantum gravity corrections described by the integral terms on the right-hand side are indeed small in the perturbation parameter since they involve the derivative of $\chi$ with respect to the gravitational variables, which are of $\ord{M^{-1}}$ due to the BO approximation \eqref{eq:Skin-condition-slow-derivatives}. Differently from the approaches in Section~\ref{sec:WKB}, here the obtained modified dynamics is unitary since the correction terms in Equation~\eqref{eq:final-kin-action} involve the conjugate momenta to the gravitational variables and the function $S_0$ which is constrained to be real from the HJ Equation \eqref{eq:Skin-HJ}. A cosmological implementation of this model can be found in \cite{bib:maniccia-montani-2021}.

\subsection{Fixing a Gaussian Reference Frame}\label{ssec:gaussian-ref-frame}
The implementation in Section~\ref{ssec:kin-action} managed to define a time parameter for the matter evolution overcoming the non-unitarity problem, however the connection between the kinematical action \eqref{eq:defSkin} and the reference system itself is not straightforward.
In this sense Kuchar later proceeded, together with Torre, to study the implementation of a term more clearly related to the reference frame~\cite{bib:kuchar-torre-1991}. In this further work, the additional term corresponds to the selection of the Gaussian reference frame $\gamma^{00} = 1$, $\gamma^{0i}=0$ reparametrized in terms of generic coordinates
\begin{linenomath}
\begin{equation}\label{eq:Sfluid-parametrized}
    S_f = \int d^4x \left[ \frac{\sqrt{-g}}{2}\, \mathcal{F} \left(g^{\alpha \beta} \partial_{\alpha}T(x) \, \partial_{\beta}T(x) -1 \right) +\sqrt{-g}\, \mathcal{F}_i \left( g^{\alpha \beta} \partial_{\alpha}T(x)\, \partial_{\beta}X^i(x) \right) \right]\,.
\end{equation}
\end{linenomath}
In Equation~\eqref{eq:Sfluid-parametrized}, $X^i(x^{\alpha}), T(x^{\alpha})$ are the Gaussian coordinates written in terms of the general $x^{\alpha}$ whose associated metric is $g_{\alpha \beta}$, and $\mathcal{F}, \mathcal{F}_i$ act as Lagrange multipliers. In this notation, $\partial_{\alpha} X^i = \partial X^i (x^{\alpha})/\partial x^{\alpha}$ and the dependence of the Gaussian coordinates on the $x^{\alpha}$ will be implied. The choice of the Gaussian coordinates is based on a straightforward implementation of fixing a reference frame (see also \cite{bib:montani-cianfrani-2008-synchronous}), while {the case of parametrized unimodular gravity is discussed in} \cite{bib:kuchar-1991-cosmoconst}, {see also}  \cite{bib:magueijo-2021} and the general parametrization process has been addressed in \cite{bib:ishamkuchar-1985}. The so-called Kuchar--Torre model is characterized by the emerging of such Gaussian reference frame as a heat-conducting fluid in the theory. This brings a source term in Einstein's equations
\begin{linenomath}
\begin{equation}\label{eq:Tmunu-fluid}
    T^{\alpha \beta} = \mathcal{F}\, \mathrm{U}^{\alpha} \mathrm{U}^{\beta} + \frac{1}{2} \left( \mathcal{F}^{\alpha}\,\mathrm{U}^{\beta} + \mathcal{F}^{\beta} \,\mathrm{U}^{\alpha} \right)\,,
\end{equation}
\end{linenomath}
being $\mathrm{U}^{\alpha} = g^{\alpha \beta} \partial_{\beta} T $ the four-velocity of the fluid, $\mathcal{F}$ its energy density, and $\mathcal{F}_{\alpha} = \mathcal{F}_i \partial_{\alpha}X^i $ its heat flow. Actually, implementing only the Gaussian time condition in \eqref{eq:Sfluid-parametrized}, the fluid reduces to an incoherent dust since $\mathcal{F}_i$ is not needed and the stress energy tensor \eqref{eq:Tmunu-fluid} reduces to the typical form $\mathcal{F}\, \mathrm{U}^{\alpha} \mathrm{U}^{\beta}$.
It is clear from Equation~\eqref{eq:Tmunu-fluid} that the fluid emerges at the classical level acting as a source term for the gravitational sector; for this reason, the fluid has to satisfy the related energy conditions in order to be physical and not ill-defined. As examined in the original work, this corresponds to the following relation
\begin{linenomath}
\begin{equation}\label{eq:KT-energy-conditions}
    \mathcal{F} \geq 2 \sqrt{\gamma^{\alpha \beta} \mathcal{F}_{\alpha} \mathcal{F}_{\beta}} \,.
\end{equation}
\end{linenomath}
However, this condition is not satisfied in principle and it is also not conserved during the evolution unless the system is closed with an additional constraint that turns the fluid to an incoherent dust and reduces \eqref{eq:KT-energy-conditions} to $\mathcal{F} \geq 0$. Thus, the energy conditions are not satisfied in the general case $\mathcal{F}, \mathcal{F}_i \neq 0$, while it is possible in the incoherent dust case $\mathcal{F}_i =0$ with some suitable initial conditions. 

In the Hamiltonian formalism, the total superspace constraints must vanish, containing the additional functions
\begin{linenomath}
\begin{gather}
    \label{eq:KTfluidH}
    H^f = W^{-1} P + W W^k P_k \, ,\\
    \label{eq:KTfluidHi}
    H_i^f = P\, \partial_i T + P_k \, \partial_i X^k\,;
\end{gather}
\end{linenomath}
where the Lagrange multipliers have been written in terms of the momenta $P$, $P_k$ conjugate to $(T, X^k)$ and the functions $W$, $W^k$ are defined as
\begin{linenomath}
\begin{gather}
    W \equiv (1- h^{jl} \partial_j T \,\partial_l T)^{-1/2} \, ,\label{eq:KTdefW}\\
    W^k \equiv h^{jl} \partial_j T \,\partial_l X^k  \,.\label{eq:KTdefWk}
\end{gather} 
\end{linenomath}
As in \eqref{eq:defHkin} and \eqref{eq:defHikin}, the momenta linearly appear in the constraints. Indeed, the authors show that defining the time derivative from the reference fluid variables, the gravity-fluid system is described by a Schr\"{o}dinger dynamics 
\begin{linenomath}
\begin{equation}\label{eq:kuchar-torre-time-def}
    i\hbar\, \partial_t \Psi = \int_{\Sigma} d^3x \frac{\delta \Psi (T,X^k,h^{jl})}{\delta T(x)} \Big|_{T=t} \Psi = \hat{\mathcal{H}} \Psi = \int_{\Sigma} d^3x \, \hat{H}^g\, \Psi .
\end{equation}
\end{linenomath}
Here, the time derivative is defined in the case of ADM foliation such that the timelike direction coincides with the Gaussian time $T$ one; both the cases in which $x^i \equiv X^i$ and $t\equiv T, x^i\equiv X^i$ are also discussed in the original paper. 

Another related approach is the work \cite{bib:brown-kuchar-1995}. There, the added sector is composed of an incoherent dust whose comoving coordinates and proper time identify a \lq \lq privileged'' reference frame which again can be used to overcome the frozen formalism issue. Furthermore, the obtained functional Schr\"{o}dinger equation is independent from the dust coordinates and a conserved inner product can be defined. However, the square-root form of the dust superHamiltonian, representing the dust scalar energy density, leads to some difficulties in implementing this definition at a WKB perturbative level. For a minisuperspace application of the Kuchar--Brown dust time, using the RPSQ and BO approximation, see \cite{bib:giesel-2009}.

The possibility to implement the same WKB and BO procedure for the model with the Gaussian reference fluid term is investigated in~\cite{bib:maniccia-montani-2022}. We start from the WDW equation 
\begin{linenomath}
\begin{equation}\label{eq:WDW-with-fluid}
    \left[ \left( -\frac{\hbar^2}{2M} \left( \nabla_g^2 + g\cdot \nabla_g \right) + M \, V\right) + (-\hbar^2 \nabla_m^2 +U) + (W^{-1} P + W W^k P_k) \right] \Psi =0 ,
\end{equation}
\end{linenomath}
and the total supermomentum constraint
\begin{linenomath}
\begin{equation}\label{eq:Hi-with-fluid}
    \left[( 2 i \hbar\, h_i\, \bar{D}\cdot \nabla_g ) - (\partial_i \phi) \nabla_m + P\, \partial_i T + P_k \, \partial_i X^k \right] \Psi =0,
\end{equation}
\end{linenomath}
for generality. The BO separation is implemented as $\Psi (h_{ij}, \phi, X^{\mu}) = \psi \left(h_{ij}\right) \chi \left( \phi, X^{\mu} ; h_{ij}\right)$, where the inclusion of the Gaussian reference frame into the fast quantum sector is backed by its materialization as a fluid \eqref{eq:Tmunu-fluid}. The WKB expansion in $M$ up to $\ord{M^{-1}}$ corresponds to the same ansatz \eqref{eq:ansatz-azione-cin} with functions $Q_n = Q_n (\phi, X^{\mu}; h_{ij})$.
The adiabatic approximation of the BO procedure gives
\begin{linenomath}
\begin{gather}
     \frac{\delta Q_n}{\delta h_{ij}} = \ord{M^{-1}},\label{eq:condiz-Qi-fluid}\\
      \frac{\langle\hat{H}^m \chi\rangle}{\langle\hat{H}^g \Psi\rangle} = \ord{M^{-1}}.
\end{gather}
\end{linenomath}
Considering again the matter backreaction to be negligible at the gravitational scale, as in Section~\ref{sec:vilenkin}, the gravitational constraints \eqref{eq:Skin-superHgrav} and \eqref{eq:Skin-supermgrav} also hold.
Expanding the system of Equations \eqref{eq:WDW-with-fluid}, \eqref{eq:Hi-with-fluid}, \eqref{eq:Skin-superHgrav} and \eqref{eq:Skin-supermgrav} with the ansatz \eqref{eq:ansatz-azione-cin}, the dynamics at the lowest order $\ord{M}$ is described by the same HJ Equation \eqref{eq:Skin-HJ} and diffeomorphism invariance of $S_0$ \eqref{eq:Skin-diffeom-S0}. The order $M^0$ describes a functional Sch\"{o}dinger dynamics with time definition
\begin{linenomath}
\begin{equation}
  \begin{split}
    \label{eq:fluid-schrod-time}
     i \hbar \dtau{\chi_0}= &\int d^3x \left[ N\left( W^{-1} \frac{\delta}{\delta T} +W W^k \frac{\delta}{\delta X^k} \right) +N^i\left( (\partial_i T)\frac{\delta}{\delta T} +(\partial_i X^k) \frac{\delta}{\delta X^k} \right) \right] \chi_0  \\
     =&\hat{\mathcal{H}}^m \chi_0\,,
    \end{split}
\end{equation}
\end{linenomath}
being $\mathcal{H}^m$ the matter Hamiltonian defined as linear combination of superHamiltonian and supermomentum functions as in \eqref{eq:Skin-schrod-time}. We stress that the time derivative in \eqref{eq:fluid-schrod-time} is defined for a generic foliation since the general coordinates are left independent from the Gaussian ones, differently from \eqref{eq:kuchar-torre-time-def}. Another key property of the model is that the fluid always emerges at the quantum level, being absent from the HJ Equation \eqref{eq:Skin-HJ}, thus it does not suffer from the energy condition problem discussed above.
Finally, the order $M^{-1}$ describes quantum gravity corrections in the form
\begin{linenomath}
\begin{equation}\label{eq:final-fluid}
   i\hbar \dtau{\chi} = \hat{\mathcal{H}}^m \chi + \int d^3 x \left[ N \nabla_g S_0 \cdot \left( -i\hbar\nabla_g \right) - 2N^i h_i \bar{D} \cdot \left( -i\hbar \nabla_g \right) \right] \chi\,,
\end{equation}
\end{linenomath}
which, also in this case, are small in the parameter $M$ and unitary due to assumption \eqref{eq:condiz-Qi-fluid} and the reality of $S_0$ from the HJ equation.

Actually, we stress that the functional forms of Equations~\eqref{eq:final-kin-action} and \eqref{eq:final-fluid} depict an analogy between the kinematical action and the Gaussian reference frame implementation. Indeed, the time definitions \eqref{eq:Skin-schrod-time} and \eqref{eq:fluid-schrod-time} can be related, if one restricts the kinematical action to the form $\partial_t y^{\mu} \rightarrow \dot{T}$ by selecting the homogeneous setting $N^i=0$ and the timelike direction $n^{\mu} = (1,\vec{0})$. Moreover, taking the Gaussian reference frame fixing with only the time condition $\mathcal{F}\neq 0, \mathcal{F}_i =0$ that is the incoherent dust, the two procedures give the same dynamics both at $\ord{M^0}$ and $\ord{M^{-1}}$. This property signals that the kinematical action is playing the role of the reference frame, acting as a fast quantum matter component and giving a preferred set of variables suitable for the construction of the time parameter. However, the parallelism is not full since the two implementations \eqref{eq:Skin-schrod-time} and \eqref{eq:fluid-schrod-time} differ between each other in the case of a generic foliation. It follows that a direct correlation between the Gaussian reference frame fixing and the kinematical action is not yet understood in the general case.

\subsection{Reference Fluid as Time in the Minisuperspace}\label{ssec:fluid-applic}

We now analyze the effects of the modifications in \eqref{eq:final-fluid} for quantum cosmology in the minisuperspace, focusing on the behaviour of the probability density during the slow-rolling phase of an isotropic universe.
Let us consider the minisuperspace reduction of the FLRW model with a free inflaton scalar field $\phi$ and a positive cosmological constant $\Lambda$ in the gravitational potential accounting for the slow-roll phase of inflation \cite{bib:maniccia-montani-2022}.
This allows us to discard spatial dependencies and restrict the general form \eqref{eq:Sfluid-parametrized} to the case of a reference frame having $g^{00} =1$, {i.e.,} imposing the reparametrized constraint $g^{\mu\nu}\partial_{\mu}T\, \partial_{\nu}T - 1 = 0$ only. In this setting, the WKB expansion in $M$ (due to the related energies being below the Planck scale) and the BO separation \eqref{eq:ansatz-azione-cin} are performed, considering a negligible backreaction from the dynamical contributions of the matter scalar field. The line element takes the simple form
\begin{linenomath}
 \begin{equation}
    	ds^2 = N^2(t) \,dt^2 - a(t)^2 \left(dx^2 + dy^2 + dz^2\right) \, , 
    	\label{RWmetric}
    \end{equation}
\end{linenomath}
in which $a$ is the cosmic scale factor, while the action corresponds to
\begin{linenomath}
\begin{equation}\label{eq:S-applic-fluid}
    	S = \int d^4x\sqrt{-g} \left\{ -\frac{1}{2\kappa}\left( R + 2\Lambda\right) + \frac{1}{2}g^{\mu\nu}\partial_{\mu}\phi\, \partial_{\nu}\phi +\frac{\mathcal{F}}{2}\left(g^{\mu\nu}\partial_{\mu}T\, \partial_{\nu}T - 1\right)\right\} \, , 
\end{equation}
\end{linenomath}
where $R = 6 \left( \frac{\Ddot{a}}{a} +\frac{\dot{a}^2}{a^2} \right)$ and the spatial Lagrange multiplier $\mathcal{F}_i$ is discarded.
Due to homogeneity $\phi = \phi(t)$, $T=T(t)$, and $\mathcal{F}=\mathcal{F}(t)$. Hence, Equation~\eqref{eq:S-applic-fluid} in the Hamiltonian formulation takes the form
\begin{linenomath}
\begin{equation}\label{eq:S-applic-fluid-RW}
    S_{RW} = \int dt \left\{ p_a\,\dot{a} + p_{\phi}\,\dot{\phi} + p_T\dot{T} - N\left( -\frac{\kappa}{12}\frac{p_a^2}{a} + \frac{\Lambda}{\kappa} a^3 + \frac{p_{\phi}^2}{2a^3} + p_T \right) \right\} \,,
\end{equation}
\end{linenomath}
where $\dot{} \equiv \partial/\partial t$ and $t$ coincides with the Gaussian time $T$ due to the constraint introduced by $\mathcal{F}$ (also $N=1$ as a consequence of the Gaussian condition on the metric). The spatial integration has been removed by considering a fiducial volume $V_0=1$. In this case, the only contribution from the reference fluid is the momentum $p_T$ present in \eqref{eq:S-applic-fluid-RW}, that is related to $\mathcal{F}$ via $	p_T = a^3 \mathcal{F} \,\dot{T}/ N$. Hence, to recover the lapse function relation $\dot{T} = N$ it must hold $p_T = \mathcal{F} a^3$.
The WDW equation gives
\begin{linenomath}
\begin{equation}
     \label{eq:fluid-appl-WDW}
        \left( \frac{\hbar^2}{48 M a}\, \partial_a^2  +4M \Lambda a^3 -\frac{\hbar^2}{2a^3} \partial_{\phi}^2  -i\hbar \, \partial_T \right) \Psi =0,
\end{equation}
\end{linenomath}
where we have considered the natural ordering with $f\cdot \nabla_g \equiv 0$.

\begin{figure}[H]
  \begin{adjustwidth}{-\extralength}{0cm}
  \centering
  \includegraphics[width=7cm]{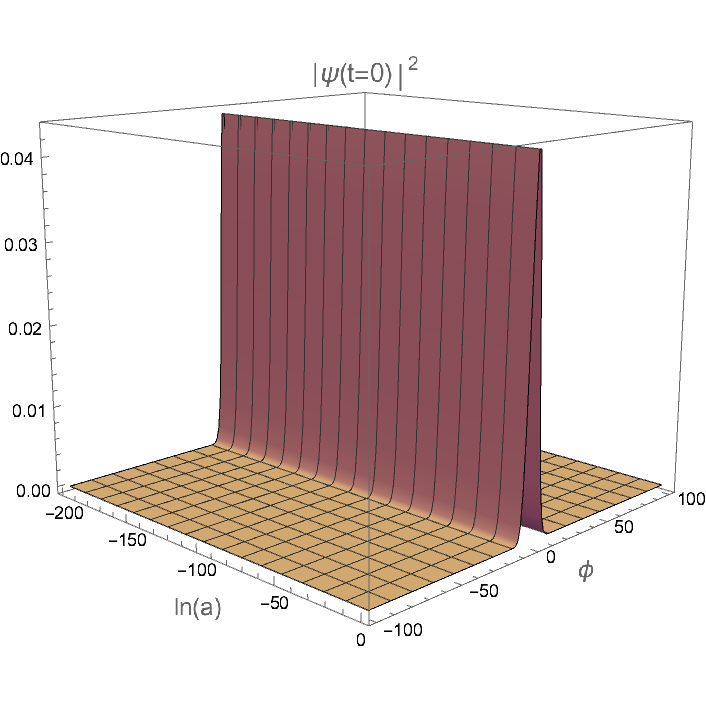}\\(\textbf{a})\\
  \includegraphics[width=7cm]{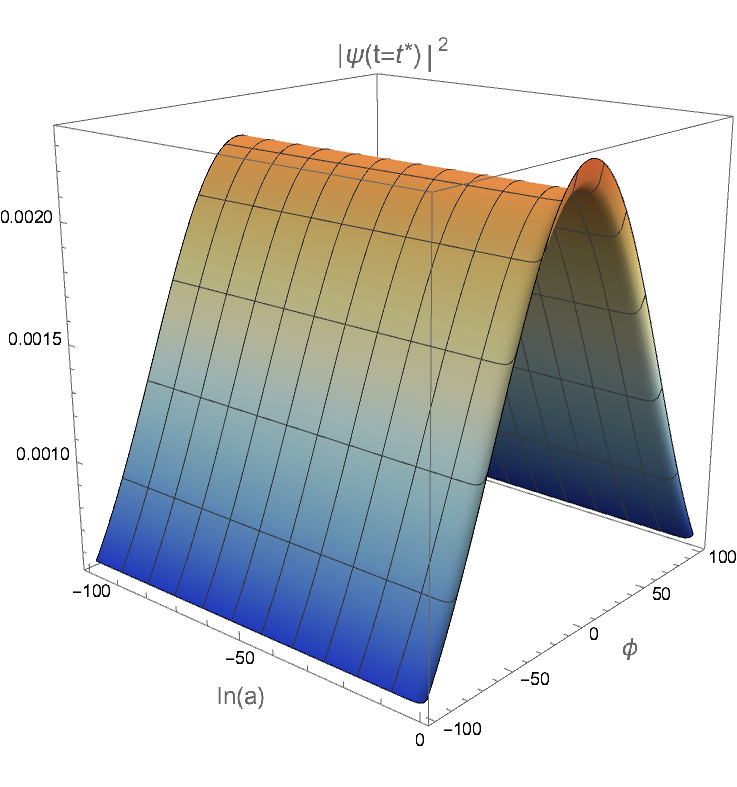}  \includegraphics[width=7cm]{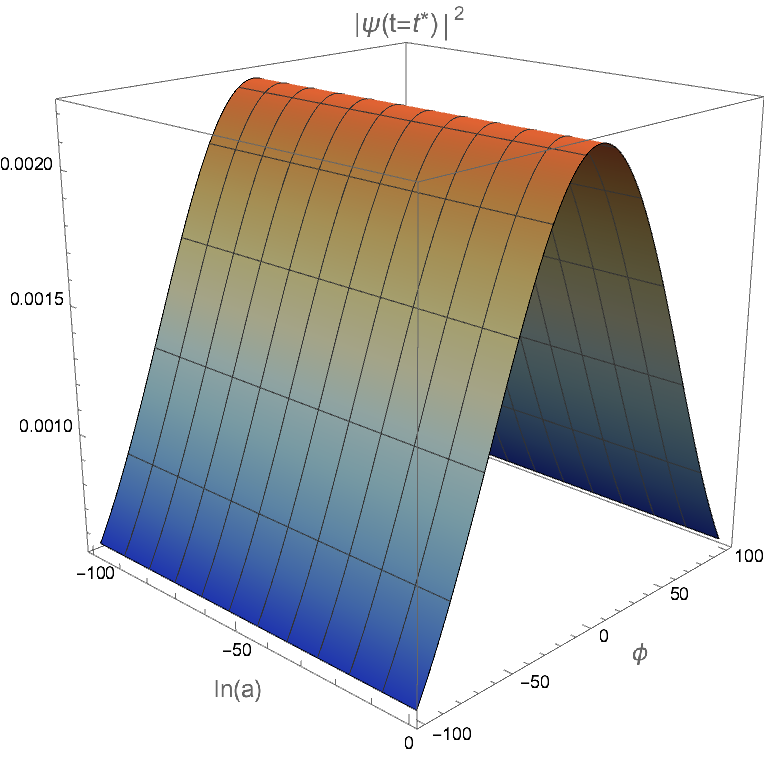}\\\hspace{0.1\linewidth}(\textbf{b})\hspace{0.4\linewidth}(\textbf{c})\\
  \end{adjustwidth}
  \caption{
  Evolution of the probability density with and without quantum gravity effects. Plot (\textbf{a}) represents the initial probability density at $\tau =0$ associated to the wavepacket \eqref{eq:applic-fluid-wavepacket} Gaussian in the variables $a, \phi$ and satisfying the condition $|p_a|\ll M^{-1}$ (we used $M=100$, $\Lambda = 10^{-2}$, $ln(a)_0=10$, $\bar{p}_{\phi}=0$, $\sigma_{\phi}=3$, $\bar{p}_a =0$, $\sigma_{a}= 2\cdot 10^{-2}$). Plots (\textbf{b}) and (\textbf{c}) show the spreading of the wave packet at later times without and with the quantum gravity effects computed in \eqref{eq:fluid-applic-eqfinal} respectively. We note that the quantum gravity corrections cause a deformation along the $a$ axis when $a$ approaches a reference value. Each wavefunction has been normalized on a suitable interval of values for the logarithm of the cosmic scale factor $\ln{(a)}$. Figures re-elaborated {from} \cite{bib:maniccia-montani-2022}.
  }
  \label{fig:plots-fluid-appl}
\end{figure}

Following the procedure of Section~\ref{sec:kin-fluid-unitarity}, Equation~\eqref{eq:fluid-appl-WDW} and the gravitational constraint~\eqref{eq:Skin-superHgrav} are expanded at each order in $M$. We emphasize that the supermomentum constraints are automatically satisfied due to homogeneity of the model, such that the respective equations are not included in the minisuperspace reduction. Numerical solutions for the gravitational functions $S_n$ are computed, selecting the ones corresponding to an expanding universe\vspace{6pt}
\begin{linenomath}
\begin{subequations}
\begin{gather}
    S_0(a) = -\frac{8\sqrt{3}}{3} \sqrt{\Lambda} \left( a^3 -a_0^3 \right)\,,\\
     S_1(a) = i\hbar \, \log \left(\frac{a}{a_0}\right) \,,\\
      S_2(a) = -\frac{\hbar^2}{24\sqrt{3} \sqrt{\Lambda}} \left( a^{-3} - a_0^{-3} \right),
\end{gather}
\end{subequations}
\end{linenomath}
where $a_0$ is an integration constant corresponding to the reference value of the scale factor at the beginning of the slow-rolling phase.
The matter sector at order $M^0$ follows the dynamics
\begin{linenomath}
\begin{equation}
    \label{eq:fluid-applic-chi0}
    -\frac{\hbar^2}{2a^3} \partial_{\phi}^2 \chi_0 = \hat{H}^m \chi_0 = i \hbar \, \frac{\partial \chi_0}{\partial T} \, ,
\end{equation}
\end{linenomath}
which is the minisuperspace reduction of \eqref{eq:fluid-schrod-time}. Solutions to \eqref{eq:fluid-applic-chi0} are, in Fourier space, the plane waves
\begin{linenomath}
\begin{equation}
    \tilde{\chi}_0 = e^{-i \hbar \frac{p_{\phi}^2}{2a^3} T},
\end{equation}
\end{linenomath}
corresponding to standard field theory evolution on curved background.
At $\ord{M^{-1}}$ such dynamics is modified by quantum gravity corrections, {such that summing with the previous order and taking into account the expansion parameter}  the equation {becomes}
\begin{linenomath}
\begin{equation}
     \label{eq:fluid-applic-eqfinal}
        \left( -\frac{\hbar^2}{2a^3} \partial_{\phi}^2 +
        i \hbar \frac{1}{24a} (\partial_a S_0) \,\partial_a \right) \chi = i \hbar \, \frac{\partial \chi}{\partial T}\,,
     \end{equation}
\end{linenomath}
for which explicit solutions can be computed in Fourier space by changing the time variable to a re-scaled time $d\tau = \frac{dT}{a^3}$ evolving with the universe volume. Then, the solution to \eqref{eq:fluid-applic-eqfinal} reads as
\begin{linenomath}
\begin{equation}
    \label{eq:fluid-applic-solchi}
    \tilde{\chi} = \exp{\left( -i\hbar \frac{ p_{\phi}^2 }{2} \tau \,+ i\, \frac{p_a \, (-\tau)^{7/3}}{7(3\Lambda)^{1/6}}  \right)} \,,
    \end{equation}
\end{linenomath}
where the smallness of the corrections is ensured by the hypothesis $|p_a| < M^{-1}$ deriving from the adiabatic approximation \eqref{eq:condiz-Qi-fluid}. As discussed before, the kinematical action implementation would describe the same modified dynamics in the minisuperspace setting. The solution \eqref{eq:fluid-applic-solchi} corresponds to a time-dependent shift in the matter energy spectrum 
\begin{linenomath}
\begin{equation}
    \label{eq:applModifiedSpectrum}
    E = E_0 + \frac{\hbar p_a (-\tau)^{7/3}}{3 (3\Lambda)^{1/6}}\, .
    \end{equation}
\end{linenomath}
To better investigate its effects, it is useful to construct an initial Gaussian wavepacket 
\begin{adjustwidth}{-\extralength}{0cm}
\begin{equation}\label{eq:applic-fluid-wavepacket}
     \chi(a, \phi, T) = \int dp_{\phi} \int dp_a \,\tilde{\chi} (p_{\phi}, p_a, T) \, \frac{1}{\sqrt{(2\pi)^{1/2}\, \sigma_a}} \exp{\left( -\frac{(p_a-\bar{p}_a)^2}{4\sigma_a^2}\right)}\, \frac{1}{\sqrt{(2\pi)^{1/2}\, \sigma_{\phi}}} \exp{\left(-\frac{(p_{\phi}-\bar{p}_{\phi})^2}{4\sigma_{\phi}^2}\right)},
\end{equation}
\end{adjustwidth}
where $\sigma_a, \bar{p}_a$ and $\sigma_{\phi}, \bar{p}_{\phi}$ describe the standard deviation and mean value of the wavepacket associated to the gravitational and matter variable respectively. The wavefunction has a small dependence on the scale factor $a$, due to the condition \eqref{eq:condiz-Qi-fluid} on $p_a$, as shown in Figure~\ref{fig:plots-fluid-appl}.

The probability density associated to \eqref{eq:applic-fluid-wavepacket} using the solution $\tilde{\chi}$ in \eqref{eq:fluid-applic-solchi} can be investigated at different values of the rescaled time $\tau$. Figure~\ref{fig:plots-fluid-appl} illustrates the effects of the time-dependent modifications \eqref{eq:fluid-applic-solchi}, which are computed for the maximum $\tau$ in the allowed domain. We stress that near the Planck scale, that is outside of this domain, the previous approximations break down and one should consider an alternative algorithm to infer the evolution of the matter dynamics.

\section{Discussion and Conclusions}\label{sec:conclusions}

We analysed different aspects concerning the separation of a system phase space into a quasi-classical part and a small quantum subsystem. We first discussed the original idea in~\cite{bib:vilenkin-1989} about the possibility to re-construct a Schr\"{o}dinger equation for the quantum variables and then we considered also the possibility to include quantum effects of the quasi-classical system into the quantum evolution of the small subsystem. 

The first analysis had the main task to show how, when applied to the mini-superspace of the Bianchi models, this approach is able to provide interesting implications on the nature of the so-called corner configuration \cite{bib:montani-chiovoloni-cascioli-2020,bib:deangelis-2020}. In particular, the possibility of a non-singular picture of the Bianchi VIII and IX dynamics, as well as for the generic cosmological solution, emerged. The crucial point was here the non-singular behavior of the Bianchi I dynamics when the variable $\beta_-$ is vanishing. Since, according to {the method in} \cite{bib:vilenkin-1989} (see also \cite{bib:montani-chiovoloni-2021}), this variable dynamics is described via the Schr\"{o}dinger equation of a time-dependent harmonic oscillator, we arrive to describe the corner dynamics via a steady classical universe over which a very small quantum anisotropy still lives. Actually, $\beta_-$ has a probability distribution peaked around its zero value and characterized by a constant small anisotropic standard deviation. 
As extended to a generic inhomogeneous cosmological solution, this picture offers an intriguing paradigm to solve the problem of the initial singularity. 

It is also an interesting achievement to have demonstrated that, comparing the Bianchi I model described in the ADM quantization procedure (also known as reduced phase space quantization \cite{bib:montani-primordialcosmology,bib:cianfrani-canonicalQuantumGravity}) with Vilenkin's formulation \cite{bib:vilenkin-1989}, the coincidence of the two approaches emerged when the quantum phase space of the anisotropic variables is sufficiently small. 
This has confirmed the consistency of the original proposal, where such an hypothesis on the quantum phase space was considered a basic statement. 

In the second part of this review, we studied the various approaches proposed in the literature to determine the possible quantum gravity corrections to quantum field theory. By other words, we consider the small quantum subsystem coinciding with matter fields, while the quasi-classical component was the background gravitational field. 

With respect to the original analysis in \cite{bib:vilenkin-1989}, the WKB procedure has been developed to the next order of approximation when quantum gravity corrections to the standard matter quantum dynamics have to arise, as in \cite{bib:kiefer-1991}. 
In particular, we re-analyzed the emerging problem that, at such further order of approximation, the Schr\"{o}dinger equation for the matter fields acquires non-unitary (non-physical) contributions.
The analyses in \cite{bib:vilenkin-1989} and in \cite{bib:kiefer-1991} have been compared, showing, on one hand, that they are essentially equivalent and, on the other hand, that some proposed solutions to the non-unitary problem \cite{bib:kiefer-2018,bib:bertoni-venturi-1996} are not consistently viable. The delicate point emerged to be the construction of a time evolution in terms of the classical dependence of the gravitational field on the label time. 
On the base of this argument, we eventually revised two different approaches in which the time coordinate belongs to the fast (matter) component of a Born--Oppenheimer scheme. 
In particular, we re-analyzed two related proposals, one based on the introduction of the so-called kinematical action \cite{bib:kuchar-1981} and one on the \lq \lq materialization'' of a fixed reference frame as a fluid, first investigated in \cite{bib:kuchar-torre-1991}.  
These formulations led to the unitary Schr\"{o}dinger equation amended for quantum gravity effects on the quantum matter dynamics. A cosmological implementation of the analysis in \cite{bib:maniccia-montani-2022} (de facto valid also for the proposal in \cite{bib:montani-digioia-maniccia-2021}) shows the consistency of the procedure and outlined some delicate questions concerning the dependence of the matter wave function on an intrinsic quantum gravity effect (there corresponding to the presence of the cosmic scale factor of the isotropic Universe), actually absent in the cosmological applications \cite{bib:brizuela-kiefer-2016-desitter,bib:brizuela-kiefer-2016-slow-roll} of the study \cite{bib:kiefer-1991}.
The present review had the scope to collect together some different efforts to amend quantum field theory for quantum gravity corrections. Our presentation elucidated that, as far as we limit our attention to the first two orders of approximation in the WKB expansion of the theory, the procedure remains consistent and it gives interesting insight on the primordial universe evolution. 
On the contrary, when the next order of approximation has been included, the one really introducing quantum gravity effects, then we have to move on a rather pioneering topic in which basic inconsistencies and intriguing proposals co-exist calling attention for further investigation in the future. 

We conclude by observing that the analysis in Sec.~\ref{sec:kin-fluid-unitarity} provides an interesting framework to search for phenomenological fingerprints of the quantum gravity corrections to quantum field theory. In particular, the determination of the primordial spectrum of the inflaton field is a natural arena to test the predictivity of such kind of reformulations toward observations of the microwave background radiation (see for instance \cite{bib:planck-results-2016,bib:cabass-divalentino-2016}). Furthermore, the analysis in Subsec.~\ref{ssec:taub} gives a significant insight on the possibility that pre-inflationary tensor perturbations survive in the later universe and can leave a trace in the B-modes of the microwave background spectrum~\cite{bib:planck-bmodes-2016}.

\vspace{6pt}
\authorcontributions{All authors contributed to all parts of the conceptual definition of the review and to the writing of the manuscript. All authors have read and agreed to the published version of the manuscript.}

\funding{This research received no external funding.}

\dataavailability{Not applicable}

\acknowledgments{G. Maniccia thanks the TAsP INFN initiative for support.}

\conflictsofinterest{The authors declare no conflict of interest.}

\begin{adjustwidth}{-\extralength}{0cm}

\reftitle{References}

\bibliography{review-final}

\end{adjustwidth}
\end{document}